\documentclass[aps,prl,reprint]{revtex4-1}
\usepackage{blindtext}
\usepackage{graphicx}
\usepackage[colorlinks,linkcolor=blue,anchorcolor=blue,citecolor=blue,urlcolor=blue]{hyperref}

\begin{document}
\title{Origin of instability in dynamic fracture}

\author{Chuang-Shi Shen}
 \email{chshshen@nwpu.edu.cn}
\affiliation{School of Aeronautics, Northwestern Polytechnical University, Xi'an 710072, China}%


\begin{abstract}
Unstable growth of cracks (rough crack surface and crack branching) in dynamic fracture has long been observed in various materials. Until now, there was no universally agreed upon explanation for these instabilities. Here, we demonstrate that: 1) Due to the non-uniform stress distribution in the cracked body and the expansion of high stress region as the crack velocity increases, a force-bearing cracked body can be treated as a temporary layer-like material (TLLM) with a ``brittle layer'' that the area of its automatically increases. 2) For this TLLM, it takes only one crack for the stress in its ``brittle layer'' to fall below a certain value at small velocities. Coupled with the asymmetry of the whole system, this results in a rough crack surface. 3) The ``brittle layer'' is large enough  when the crack velocity reaches a certain value. Two cracks must be formed, or else the stress cannot be reduced below the certain value, resulting in crack branching. Crack propagation can be compared to cars traveling on a road: the high stress region in the cracked body provides a ``road'' to let cracks ``pass .''
\end{abstract}

\maketitle

\begin{figure*}
	\includegraphics[scale=1.05]{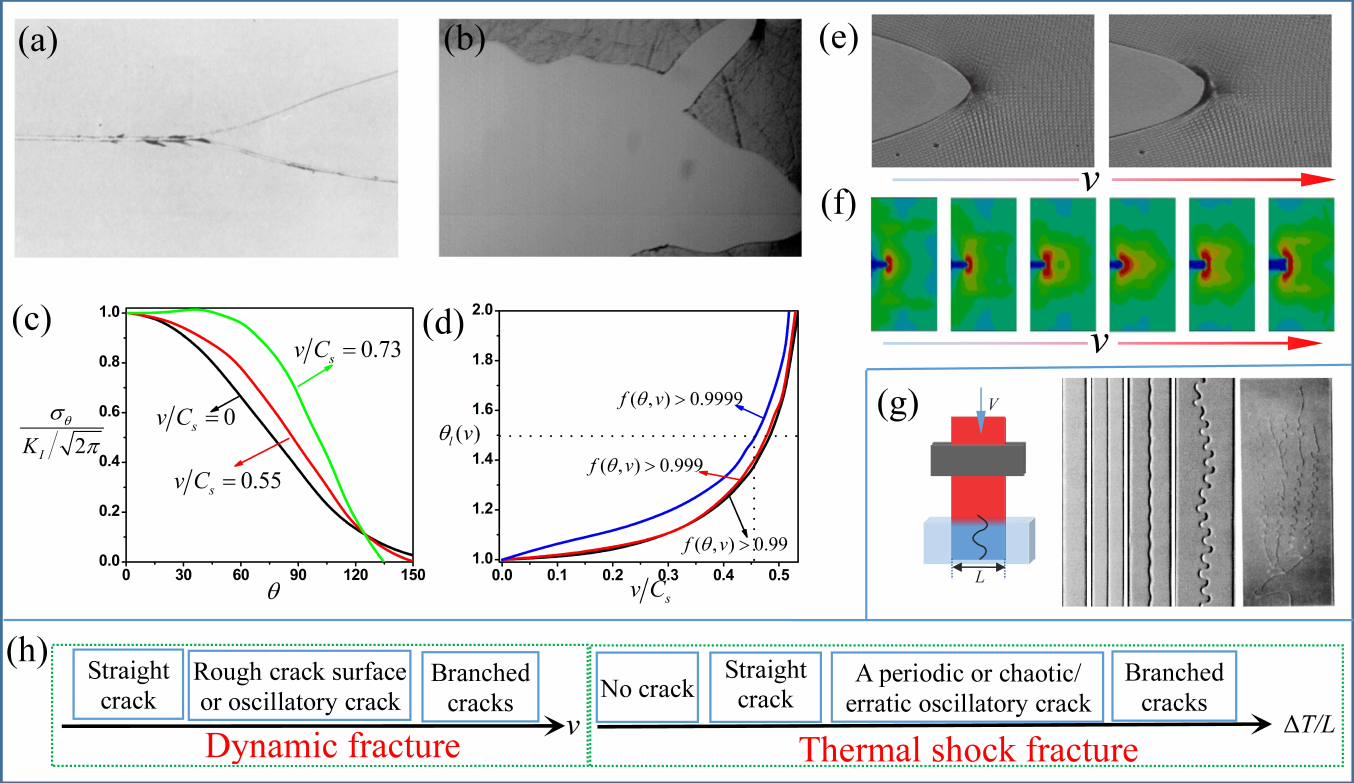}
	\caption{\label{fig1} 
		(color online) Unstable growth of cracks. Crack branching in (a) hard materials (Homalite-100) \cite{Ramulu1985} and (b) soft materials (polyacrylamide gels) \cite{ Lubomirsky2018}. (c) change of hoop stress field \cite{Yoffe1951}. (d) normalized length of high stress region. (e) lager deformation region increases with crack velocity \cite{Goldman2015}. (f) evolution of principal stress during crack propagation \cite{Pereira2017}. (g) thermal shock fracture \cite{Yuse1993,Ronsin1998}. (h) similarity between dynamic fracture and thermal shock fracture. }
\end{figure*}

\textbf{Introduction.} Dynamic fracture is of fundamental and practical importance. It has attracted widespread attention from engineers as well as physicists, and has been frequently studied for the past 80 years \cite{Schardin,Bowden1967,Ramulu1985,Fineberg1999,Cox2005,Marder1991,Langer1993,Ching1996,Adda-Bedia2004,Adda-Bedia2013,Brener1998,Lund1996,Bouchbinder2007,Deegan2002,Petersan2004,Persson1998,Holian1997,Hauch1999,Sharon1999,Stroh1957,sih1972,Quinn2019}. Once the brittle material begins to break (Mode I), and if the stretch is sufficient, the crack velocity $v$ accelerates toward a material-dependent critical velocity ${v_c}$ (This was first discovered by Schardin and Struth in 1937 \cite{Schardin}, ${v_c} \approx 0.33{C_R} \sim 0.66{C_R}$ for hard materials \cite{Quinn2019,Ravi-Chandar2004,Ravi-Chandar1984}, smaller than the theoretical limiting velocity ${C_R}$ \cite{Stroh1957,Freund1990}. ${v_c} \approx 0.9{C_S}$ for two-dimensional (2D) soft materials \cite{Goldman2010,Chen2017,Lubomirsky2018}. ${v_c} \approx 0.34{C_R}$ for three-dimensional (3D) soft materials \cite{Livne2005}. $C_R$, $C_S$ and $C_L$ represent Rayleigh, shear and longitudinal wave velocities, respectively). Then, the velocity starts to oscillate, correlating with the development of sound emission and crack front waves \cite{Gross1993,Boudet1998,Sharon2001,Sharon2002,Massy2018}. The critical velocity is usually accompanied by crack branching, as shown in Fig. \ref{fig1}(a, b). In 1996, Sharon et al. \cite{Sharon1996} concluded that: ``due to the energy is diverted to the branching cracks cause to the main crack to slow, when the branching crack dies, the energy is re-diverted to the main crack which consequently accelerates until the occurrence of the next branching event.'' This means that crack branching sets the critical velocity \cite{Sharon1995,Sharon1996,Kobayashi1972}. Below the critical velocity, an increase in crack surface roughness is observed in hard materials \cite{Fineberg1991,Fineberg1992,Sharon199601,Hull,Ravi-Chandar1984II,Ravi1997,Cramer2000,Scheibert2010,Claudia2012}, and oscillatory cracks (oscillation amplitude increases with crack velocity) are formed in 2D soft materials \cite{Chen2017, Lubomirsky2018,Livne2007,Goldman2012}. Linear elastic fracture mechanics \cite{Inglis1913,Griffith1921} falls short of explaining these phenomena \cite{Fineberg2015,Bouchbinder2014,Bouchbinder2010}. Various hypotheses existing in the literature (stress field rearrangement \cite{Yoffe1951}, interaction of microcracks \cite{Ravi-Chandar1984}, shear waves \cite{Bonamy2003}, crack front waves \cite{Bouchaud2002}, acoustic waves \cite{Massy2018}, physical discontinuity or interference of bouncing stress wave \cite{Pereira2017}, stress waves piling-up in front of the crack tip \cite{Bobaru2015}, near-tip elastic nonlinearity \cite{Chen2017,Lubomirsky2018,Goldman2012,Livne2008,Bouchbinder2008,Bouchbinder2009,livne2010}, hyperelasticity \cite{Buehler2003,Buehler2006}, anisotropic elasticity \cite{Abraham1996}, local phase transition \cite{Buehler2007,Boulbitch2011}, shear perturbation \cite{Goldman2015}, shear stress \cite{Adda-Bedia1996,Adda-Bedia1999} and  thermal noise \cite{Sander1999}) were suggested to be the possible sources for dynamic fracture instability. Bouchbinder et al. \cite{Bouchbinder2014} stated in review article: ``quite like the dynamics of crack tip zone could play an important role in unraveling the physical mechanism driving \textit{other} instabilities of rapid cracks.'' Despite dynamic fracture being reproduced very well by plenty of number simulations (lattice models \cite{Marder1993,Marder1995}, peridynamics \cite{Bobaru2015}, molecular dynamics \cite{Abraham1994,Nakano1995,Zhou1996,Omeltchenko1997,Abraham1997,Holland1998,Yamakov2005,Swadener2002,Abraham2003,Buehler200601,Atrash2011}, phase field models \cite{Chen2017,Lubomirsky2018,Pereira2017,Aranson2000,Karma2004,Henry2004,Bleyer2017,Spatschek2006}, cohesive zone models \cite{Xu1994,Nguyen2004,Elmukashfi2012} and extended finite element method \cite{Belytschko2003}), a clear physical picture of dynamic crack propagation still remains elusive. The actual mechanism that triggers the instability in dynamic fracture has not been clearly identified \cite{Lubomirsky2018,Fineberg2015,Bouchbinder2014,Goldman2015,Hellemans1998}.

Unstable growth of cracks not only occurs in dynamic fracture, but also occurs in quasi-static fracture, thermal shock fracture \cite{Yuse1993}. As shown in Fig. \ref{fig1}(g), when a hot thin glass strip is dipped into cold water at a constant velocity \cite{Yuse1993,Ronsin1995,Yuse1997,Ronsin1998,Yang2001,Xu2018}, an interesting transition occurs between no crack, a straight crack, a periodic or erratic oscillatory crack and two or more branched cracks, mainly depends on width $L$ \cite{Ronsin1995} and temperature difference $\Delta T$ \cite{Yuse1997}. There is a certain similarity between dynamic fracture and thermal shock fracture as shown in Fig. \ref{fig1}(h). Thus, the physical mechanism behind them might be one and the same.

In this study, we demonstrate, once a force-bearing cracked body is considered as a TLLM, the instability in dynamic fracture can be explained. 

\textbf{Three criteria for fracture.} The principles governing the fracture can be summarized as the following three criteria: 1) crack initiates in a body where stress equals to some critical value (the first strength theory); 2) stress is released as a result of fracture so that it falls below critical value in other parts of the body; and 3) energy consumed due to the forming of cracks is minimized (the principle of least action). The second criterion promotes the generation of more cracks, while the third criterion suggests the opposite. The compromise between the two competing criteria determines the number of cracks.

\textbf{High stress region in crack tip.} On the assumption of small deformation, Yoffe \cite{Yoffe1951} proposed a model in which a finite length crack moves along the direction of the main crack at constant velocity in a linear elastic solid, the hoop stress field in the crack tip is as follows \cite{Freund1990,Yoffe1951}
\begin{equation}
\label{eq1}
\begin{array}{l}
{\sigma _\theta }{\rm{ = }}\frac{{{K_I}}}{{\sqrt {2\pi} }}f\left( {\theta ,v} \right)\\
f\left( {\theta ,v} \right) = C({C_1}{\sin ^2}\theta  + {C_2}{\cos ^2}\theta  - 2{C_3}\sin \theta \cos \theta )\\
C = \frac{{1 + a_2^2}}{{4{a_1}{a_2} - {{(1 + a_2^2)}^2}}}\\
{C_1} = (1 + 2a_1^2 - a_2^2)\frac{{\cos \left( {{{{\theta _1}} \mathord{\left/
					{\vphantom {{{\theta _1}} 2}} \right.
					\kern-\nulldelimiterspace} 2}} \right)}}{{\sqrt {{r_1}} }} + \frac{{4{a_1}{a_2}}}{{1 + a_2^2}}\frac{{\cos \left( {{{{\theta _2}} \mathord{\left/
					{\vphantom {{{\theta _2}} 2}} \right.
					\kern-\nulldelimiterspace} 2}} \right)}}{{\sqrt {{r_2}} }}\\
{C_2} =  - (1 - a_2^2)\frac{{\cos \left( {{{{\theta _1}} \mathord{\left/
					{\vphantom {{{\theta _1}} 2}} \right.
					\kern-\nulldelimiterspace} 2}} \right)}}{{\sqrt {{r_1}} }} + \frac{{4{a_1}{a_2}}}{{1 + a_2^2}}\frac{{\cos \left( {{{{\theta _2}} \mathord{\left/
					{\vphantom {{{\theta _2}} 2}} \right.
					\kern-\nulldelimiterspace} 2}} \right)}}{{\sqrt {{r_2}} }}\\
{C_3} = 2{a_1}(\frac{{\sin \left( {{{{\theta _1}} \mathord{\left/
					{\vphantom {{{\theta _1}} 2}} \right.
					\kern-\nulldelimiterspace} 2}} \right)}}{{\sqrt {{r_1}} }} - \frac{{\sin \left( {{{{\theta _2}} \mathord{\left/
					{\vphantom {{{\theta _2}} 2}} \right.
					\kern-\nulldelimiterspace} 2}} \right)}}{{\sqrt {{r_2}} }})\\
{a_1} = \sqrt {1 - {{(v/{C_L})}^2}} \begin{array}{*{20}{c}}
{}&{{a_2} = \sqrt {1 - {{(v/{C_S})}^2}} }
\end{array}\\
{r_1} = \frac{{r\cos (\theta )}}{{\cos \left[ {\arctan ({a_1}\tan \theta )} \right]}}\begin{array}{*{20}{c}}
{}&{{r_1} = \frac{{r\cos (\theta )}}{{\cos \left[ {\arctan ({a_2}\tan \theta )} \right]}}}
\end{array}
\end{array}
\end{equation}
where ${{K_I}}$ is the stress intensity factor; $\theta $ and $r$ are polar-coordinates centered at the crack tip with $\theta $ measured counterclockwise from the growth direction. Fig. \ref{fig1}(c) shows the change of hoop stress ${\sigma _\theta }$ with crack velocity.

The normalized length of the high stress region is as follows
\begin{equation}
\label{eq2}
{\theta _l}(v) = {{f(\theta ,v)} \mathord{\left/
		{\vphantom {{f(\theta ,v)} {f(\theta ,0)}}} \right.
		\kern-\nulldelimiterspace} {f(\theta ,0)}} \sim {v^2}\
\end{equation}

As shown in Fig. \ref{fig1}(d), the high stress region enlarges as the crack velocity increases. The normalized length roughly doubled when the crack velocity increased to ${0.53C_S}$. Similar results have been observed in experiments \cite{Goldman2015,Sundaram2018} and through numerical simulations. \cite{Pereira2017,Buehler2006,Abraham1997,Abraham2003,Nguyen2004,Elmukashfi2012}. For instance, Goldman et al. \cite{Goldman2015} investigated the dynamic fracture of a soft material (polyacrylamide gels) and demonstrated that the larger deformation region in the crack tip changes its shapes as the crack velocity increases, as shown in Fig. \ref{fig1}(e). Sundaram and Tippur \cite{Sundaram2018} investigated the dynamic crack initiation, growth and branching of a hard material (soda-lime glass) and exhibited that the maximum distance ahead of the crack tip (at which the stress becomes equal to the tensile strength of a material) increases with the crack velocity. Immediately before branching, the simulation results carried out by Pereira et. al \cite{Pereira2017} shows  that the high stress region at the crack tip approximately doubled, as shown in Fig. \ref{fig1}(f).  The high stress region in the crack tip actually expends with the increased in external load even before fracture (when there is no load applied, there is no high stress region) \cite{Inglis1913,Griffith1921}. This trend will continue in dynamic fracture.

\begin{figure*}
	\includegraphics[scale=0.9]{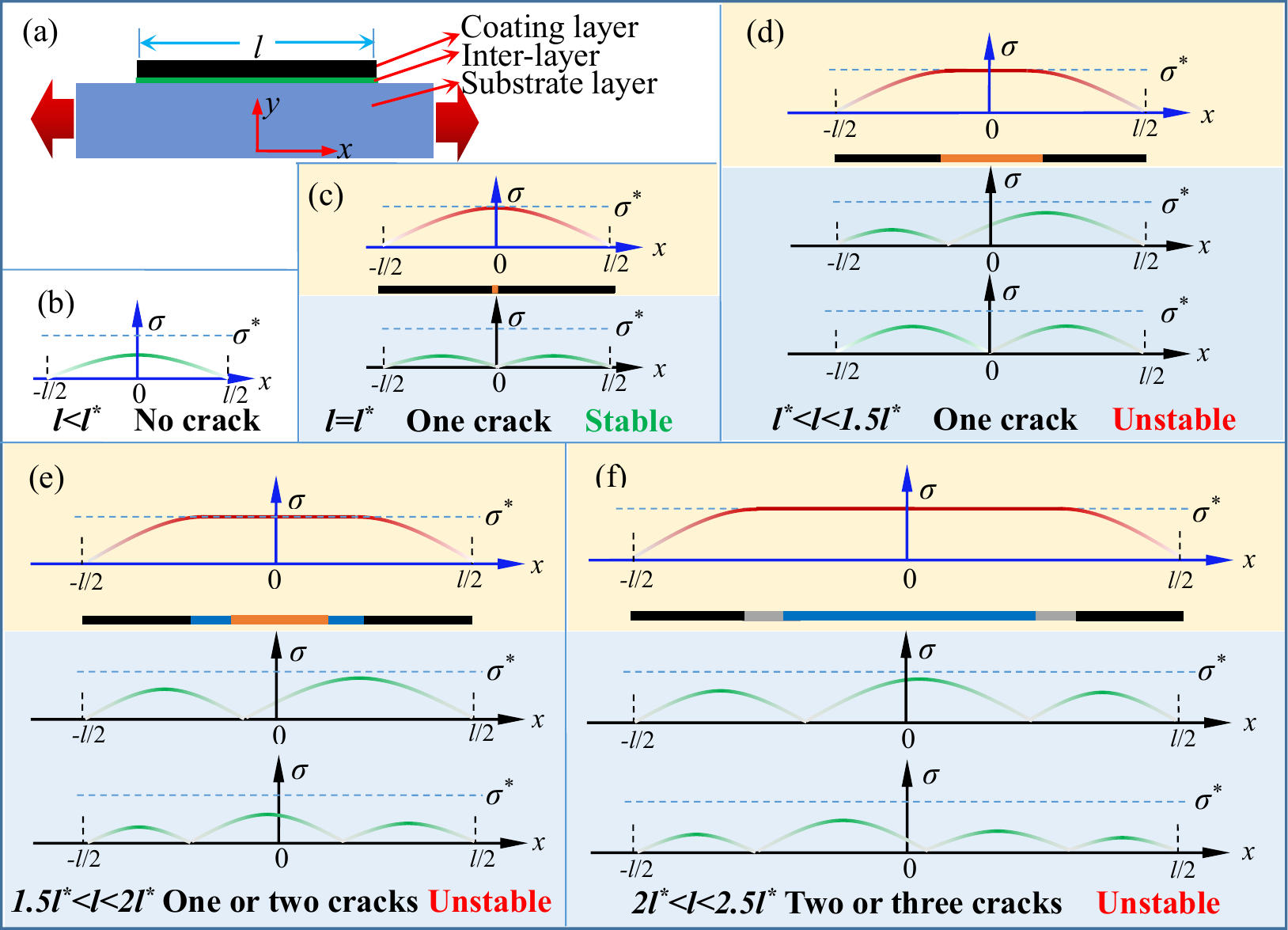}
	\caption{\label{fig2}  
		(color online) The fracture of layered materials. (a) Schematic diagram of a brittle coating layer bond to a substrate layer. (b-f) Schematic diagram of the stress distribution in coating layer before fracture (red line) and after fracture (green line).}
\end{figure*}

\textbf{Fracture of layered materials.} Leaving the dynamic fracture aside for a moment, consider the fracture of layered materials. As shown in Fig. \ref{fig2}(a), a brittle coating layer (thickness ${h_1}$, Young’s modulus ${E_1}$, Poisson’s ratio ${v_1}$, tension strength ${\sigma ^ * }$) is fully bonded to a substrate layer (thickness ${h_2}$, Young’s modulus ${E_2}$, Poisson’s ratio ${v_2}$), ${h_1} \ll {h_2}$, and subjected to a uniform tensile load (its tensile strain is equal to $\varepsilon $) in the substrate. Forces will be transferred to the coating layer through inter-layer (thickness ${h_3}$, shear modulus ${G_3}$), ${h_3} \ll {h_2}$. As shown in Fig. \ref{fig2}(c), the maximum stress appears in the midpoint of the coating layer. When this stress is equal to the tensile strength, it corresponds to the critical length ${l^ * } = 2\sqrt {{{{E_1}{h_1}{h_3}} \mathord{\left/{\vphantom {{{E_1}} {{G_3}}}} \right.\kern-\nulldelimiterspace} {{G_3}}}} \cosh^{ - 1} \left[ {{1 \mathord{\left/{\vphantom {1 {\left( {1 - {{{\sigma ^ * }} \mathord{\left/{\vphantom {{{\sigma ^ * }} {{E_1}\varepsilon }}} \right.\kern-\nulldelimiterspace} {{E_1}\varepsilon }}} \right)}}} \right.\kern-\nulldelimiterspace} {\left( {1 - {{{\sigma ^ * }} \mathord{\left/{\vphantom {{{\sigma ^ * }} {{E_1}\varepsilon }}} \right.\kern-\nulldelimiterspace} {{E_1}\varepsilon }}} \right)}}} \right]$ \cite{Cox1952,Mcguigan2003}. According to the second fracture criterion, the segment length after fracture must be less than ${l^ * }$  . Due to the maximum stress only appearing in the midpoint of the coating layer when $l = {l^ * }$, the fracture location is unique and the fracture is stable. The coating layer will not fracture when $l < {l^ * }$, because the maximum stress is less than the material strength as shown in Fig. \ref{fig2}(b). The stresses are equal in the middle segment of the coating layer before fracture when $l > {l^ * }$, as shown in Fig. \ref{fig2}(d-f). Together with the defects in materials, leads to the fracture location not being unique compared with $l = {l^ * }$, resulting in an unstable fracture. The number of cracks and fracture locations in the coating layer are closely related to the length of the coating layer. 

Therefore, for layered or layer-like materials, two cracks may be formed when 
\begin{equation}
\label{eq3}
1.5{l^ * } < l < 2{l^ * }
\end{equation}
The number of cracks shall not be less than two if
\begin{equation}
\label{eq4}
l > 2{l^ * }
\end{equation}

\begin{figure*}
	\includegraphics[scale=1]{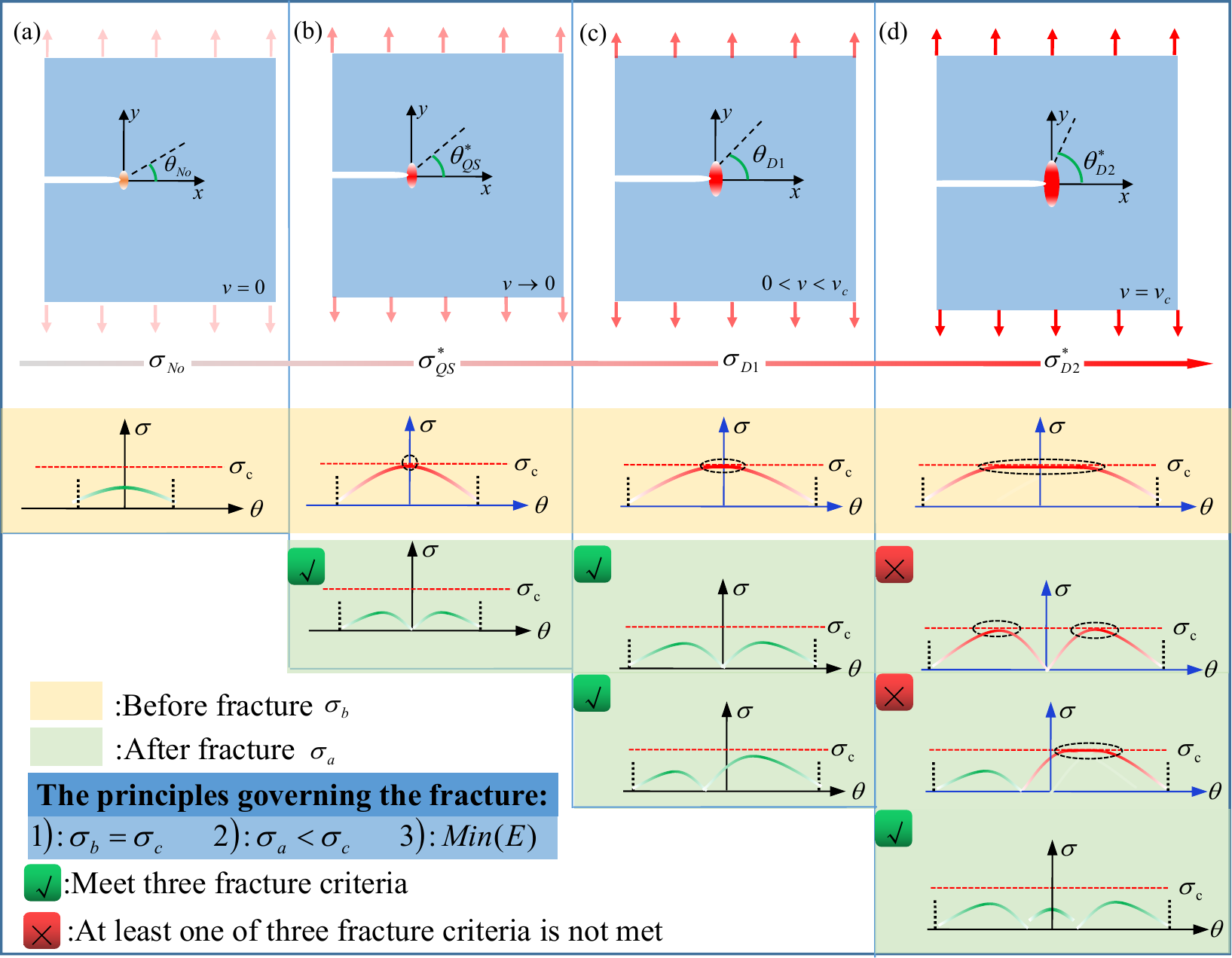}
	\caption{\label{fig3}
		(color online). Schematic diagram of dynamic fracture.  The area of the high stress region in the cracked body is positively correlated with the crack velocity. The options for fracturing the high stress region are shown in light green zone. Feasible ($\surd$)  and infeasible ($\times$) schemes.}
\end{figure*}

\textbf{Instability in dynamic fracture.} Consider a cracked body subjected to external loads, as shown in Fig. \ref{fig3}. The stress is high around the crack tip, and low everywhere else in the body. Treat the high stress region as the brittle layer in layered materials in Fig. \ref{fig2}(a), and the low stress region as the substrate layer.  Thus, a force-bearing cracked body can be considered as a TLLM. This TLLM is formed due to the presence of cracks.  Other TLLMs (cracks induced by desiccation \cite{Jeremy2019}, craquelure in ceramics \cite{Bohn2005}, thermal shock crack \cite{Song2010,Aydin1988}, frozen impacted drop \cite{Ghabache2016} and broken windows \cite{Vandenberghe2013}) are shaped because a part of the body is subjected to external loads at some point, while the other parts are not. Similar to a force-bearing layered material, the stress distribution is non-uniform. Based on the theoretical \cite{Yoffe1951,Freund1990}, experimental \cite{ Goldman2015,Sundaram2018} and numerical \cite{Pereira2017,Buehler2006,Abraham1997,Abraham2003,Nguyen2004,Elmukashfi2012} studies, the spatial distribution of crack tip stress changes shape and flattens with increase in crack velocity. This means that the length of the ``brittle layer'' in this TLLM automatically increases.
 
 For every crack velocity, let the time be frozen in this moment, treat the cracked body as a TLLM and analyze how the ``brittle layer'' in it breaks to meet the three fracture criteria under the current stress state (the whole fracture process of a cracked body is the fracture of a series of TLLMs). As shown in Fig. \ref{fig3}(a), when the external load is small, the stress in angle of ${0^ \circ }$ is smaller than the critical stress ${\sigma _c}$, the crack will not propagate. With the increase in load, the stress in angle of ${0^ \circ }$ will be equal to the critical stress ${\sigma _c}$ at a certain load. In this condition (quasi-static fracture, a specific case of fracture), there is only one way to break the ``brittle layer'', that the crack can only propagate along the angle of ${0^ \circ }$, it is stable, as shown in Fig. \ref{fig3}(b). The crack velocity will accelerate if the external load further increases, resulting in there exists equal stresses over a wide region of the crack tip as shown in Fig. \ref{fig3}(c, d), leads to unstable crack growth. For small velocities, one crack is enough to meet the second and third fracture criteria. Due to the existence of material defects or tiny vibration due to applied loads, means the whole system is asymmetric, this leads to the difficulty in maintaining the fracture position in the middle, as shown in Fig. \ref{fig3}(c), results in the crack surface becoming rough, and the roughness increases due to the enlargement of the high stress region with crack velocity. As shown in Fig. \ref{fig3}(d), when the ``brittle layer'' exceeds a certain value $\theta _{D2}^ * $, one crack will not be able to meet the second fracture criteria, two cracks must be formed, crack branching occurs. 

For elastic materials, combining Fig. \ref{fig1}(d) with Eqs. (\ref{eq3}) and (\ref{eq4}), there is a possibility (the faster the speed, the higher the probability) that two cracks are formed at crack velocity between $0.45{C_s}\sim0.53{C_s}$, two cracks have to be formed when the velocity approximately  exceeds $0.53{C_s}$, this is close to the critical velocity measured in experiments $ 0.33{C_R} \sim 0.66{C_R}$  \cite{Quinn2019,Ravi-Chandar2004,Ravi-Chandar1984}. For other materials (plastic, hyperelastic, viscoelastic, etc), it is difficult to obtain an analytical solution of the stress field in the crack tip for its quasi-static fracture, let alone the dynamic fracture. Therefore, objective to quantitatively study these materials, numerical methods may be the best practical option \cite{Marder1993,Marder1995,Abraham1994,Nakano1995,Zhou1996,Omeltchenko1997,Abraham1997,Holland1998,Yamakov2005,Swadener2002,Abraham2003,Buehler200601,Atrash2011,Chen2017,Lubomirsky2018,Pereira2017,Aranson2000,Karma2004,Henry2004,Bleyer2017,Spatschek2006,Xu1994,Nguyen2004,Elmukashfi2012,Belytschko2003}.

Regarding the crack pattern formed by dipping a hot thin glass strip into cold water \cite{Yuse1993,Ronsin1995,Yuse1997,Ronsin1998,Yang2001,Xu2018}, which a two layer-like structure  (high and low temperature) is formed at the instant of insertion. Consequently, the physical mechanism as shown in Fig. \ref{fig2} drives the rest of the process.

The instability in crack propagation is similar to a vehicle-traveling phenomenon. Assume that there are two cars on a rough road. If the road is too narrow, cars can not pass through. If the road width is exactly equal to the car width, the car trajectory is a straight line. If the road is slightly wider, less than twice the car width, the trajectory might be a sinusoidal curve. If the road is wider than twice the car width, two cars may pass side by side. For fracture mechanics, external loads building a ``road'' (high stress region) in the crack tip to allow cracks to ``pass'' (propagate).

\textbf{Conclusion.} The entire scenario of this study follows a typical Aristotelian modal syllogismm: 1) The fracture of layered materials is unstable due to the three fracture criteria must be met, 2) A force-bearing cracked body can be considered as a temporary layer-like material with a high stress layer that the length of its increases as crack velocity increases, 3) Thus, for the same reason, the growth of cracks in dynamic fracture is unstable. This study mainly discusses the origin of unstable growth of cracks in 2D media. Future research should address the instability (branch lines, topological defects) in 3D media \cite{Livne2005,Livne2007,Sharon2002,Baumberger2008,Kolvin2015,Kolvin2017a,Kolvin2017b}.

\bibliographystyle{apsrev4-1} 

\begin{thebibliography}{113}%
\makeatletter
\providecommand \@ifxundefined [1]{%
 \@ifx{#1\undefined}
}%
\providecommand \@ifnum [1]{%
 \ifnum #1\expandafter \@firstoftwo
 \else \expandafter \@secondoftwo
 \fi
}%
\providecommand \@ifx [1]{%
 \ifx #1\expandafter \@firstoftwo
 \else \expandafter \@secondoftwo
 \fi
}%
\providecommand \natexlab [1]{#1}%
\providecommand \enquote  [1]{``#1''}%
\providecommand \bibnamefont  [1]{#1}%
\providecommand \bibfnamefont [1]{#1}%
\providecommand \citenamefont [1]{#1}%
\providecommand \href@noop [0]{\@secondoftwo}%
\providecommand \href [0]{\begingroup \@sanitize@url \@href}%
\providecommand \@href[1]{\@@startlink{#1}\@@href}%
\providecommand \@@href[1]{\endgroup#1\@@endlink}%
\providecommand \@sanitize@url [0]{\catcode `\\12\catcode `\$12\catcode
  `\&12\catcode `\#12\catcode `\^12\catcode `\_12\catcode `\%12\relax}%
\providecommand \@@startlink[1]{}%
\providecommand \@@endlink[0]{}%
\providecommand \url  [0]{\begingroup\@sanitize@url \@url }%
\providecommand \@url [1]{\endgroup\@href {#1}{\urlprefix }}%
\providecommand \urlprefix  [0]{URL }%
\providecommand \Eprint [0]{\href }%
\providecommand \doibase [0]{http://dx.doi.org/}%
\providecommand \selectlanguage [0]{\@gobble}%
\providecommand \bibinfo  [0]{\@secondoftwo}%
\providecommand \bibfield  [0]{\@secondoftwo}%
\providecommand \translation [1]{[#1]}%
\providecommand \BibitemOpen [0]{}%
\providecommand \bibitemStop [0]{}%
\providecommand \bibitemNoStop [0]{.\EOS\space}%
\providecommand \EOS [0]{\spacefactor3000\relax}%
\providecommand \BibitemShut  [1]{\csname bibitem#1\endcsname}%
\let\auto@bib@innerbib\@empty
\bibitem [{\citenamefont {Schardin}(1959)}]{Schardin}%
  \BibitemOpen
  \bibfield  {author} {\bibinfo {author} {\bibfnamefont {H.}~\bibnamefont
  {Schardin}},\ }\href@noop {} {\emph {\bibinfo {title} {Velocity effects in
  fracture}}}\ (\bibinfo  {publisher} {In: Averbach BL, Felbeck DK, Thomas DA
  (eds) Fracture. Wiley, New York},\ \bibinfo {year} {1959})\ pp.\ \bibinfo
  {pages} {297--330}\BibitemShut {NoStop}%
\bibitem [{\citenamefont {Bowden}\ \emph {et~al.}(1967)\citenamefont {Bowden},
  \citenamefont {Brunton}, \citenamefont {Field},\ and\ \citenamefont
  {Heyes}}]{Bowden1967}%
  \BibitemOpen
  \bibfield  {author} {\bibinfo {author} {\bibfnamefont {F.~P.}\ \bibnamefont
  {Bowden}}, \bibinfo {author} {\bibfnamefont {J.~H.}\ \bibnamefont {Brunton}},
  \bibinfo {author} {\bibfnamefont {J.~E.}\ \bibnamefont {Field}}, \ and\
  \bibinfo {author} {\bibfnamefont {A.~D.}\ \bibnamefont {Heyes}},\ }\href
  {\doibase 10.1038/216038a0} {\bibfield  {journal} {\bibinfo  {journal}
  {Nature.}\ }\textbf {\bibinfo {volume} {216}},\ \bibinfo {pages} {38}
  (\bibinfo {year} {1967})}\BibitemShut {NoStop}%
\bibitem [{\citenamefont {Ramulu}\ and\ \citenamefont
  {Kobayashi}(1985)}]{Ramulu1985}%
  \BibitemOpen
  \bibfield  {author} {\bibinfo {author} {\bibfnamefont {M.}~\bibnamefont
  {Ramulu}}\ and\ \bibinfo {author} {\bibfnamefont {A.~S.}\ \bibnamefont
  {Kobayashi}},\ }\href {\doibase 10.1007/BF00017967} {\bibfield  {journal}
  {\bibinfo  {journal} {Int. J. Fracture.}\ }\textbf {\bibinfo {volume} {27}},\
  \bibinfo {pages} {187} (\bibinfo {year} {1985})}\BibitemShut {NoStop}%
\bibitem [{\citenamefont {Fineberg}\ and\ \citenamefont
  {Marder}(1999)}]{Fineberg1999}%
  \BibitemOpen
  \bibfield  {author} {\bibinfo {author} {\bibfnamefont {J.}~\bibnamefont
  {Fineberg}}\ and\ \bibinfo {author} {\bibfnamefont {M.}~\bibnamefont
  {Marder}},\ }\href {\doibase 10.1016/s0370-1573(98)00085-4} {\bibfield
  {journal} {\bibinfo  {journal} {Phys. Rep.}\ }\textbf {\bibinfo {volume}
  {313}},\ \bibinfo {pages} {1} (\bibinfo {year} {1999})}\BibitemShut {NoStop}%
\bibitem [{\citenamefont {Cox}\ \emph {et~al.}(2005)\citenamefont {Cox},
  \citenamefont {Gao}, \citenamefont {Gross},\ and\ \citenamefont
  {Rittel}}]{Cox2005}%
  \BibitemOpen
  \bibfield  {author} {\bibinfo {author} {\bibfnamefont {B.~N.}\ \bibnamefont
  {Cox}}, \bibinfo {author} {\bibfnamefont {H.}~\bibnamefont {Gao}}, \bibinfo
  {author} {\bibfnamefont {D.}~\bibnamefont {Gross}}, \ and\ \bibinfo {author}
  {\bibfnamefont {D.}~\bibnamefont {Rittel}},\ }\href {\doibase
  10.1016/j.jmps.2004.09.002} {\bibfield  {journal} {\bibinfo  {journal} {J.
  Mech. Phys. Solids.}\ }\textbf {\bibinfo {volume} {53}},\ \bibinfo {pages}
  {565} (\bibinfo {year} {2005})}\BibitemShut {NoStop}%
\bibitem [{\citenamefont {Marder}(1991)}]{Marder1991}%
  \BibitemOpen
  \bibfield  {author} {\bibinfo {author} {\bibfnamefont {M.}~\bibnamefont
  {Marder}},\ }\href {\doibase 10.1103/PhysRevLett.66.2484} {\bibfield
  {journal} {\bibinfo  {journal} {Phys. Rev. Lett.}\ }\textbf {\bibinfo
  {volume} {66}},\ \bibinfo {pages} {2484} (\bibinfo {year}
  {1991})}\BibitemShut {NoStop}%
\bibitem [{\citenamefont {Langer}(1993)}]{Langer1993}%
  \BibitemOpen
  \bibfield  {author} {\bibinfo {author} {\bibfnamefont {J.~S.}\ \bibnamefont
  {Langer}},\ }\href {\doibase 10.1103/PhysRevLett.70.3592} {\bibfield
  {journal} {\bibinfo  {journal} {Phys. Rev. Lett.}\ }\textbf {\bibinfo
  {volume} {70}},\ \bibinfo {pages} {3592} (\bibinfo {year}
  {1993})}\BibitemShut {NoStop}%
\bibitem [{\citenamefont {Ching}\ \emph {et~al.}(1996)\citenamefont {Ching},
  \citenamefont {Langer},\ and\ \citenamefont {Nakanishi}}]{Ching1996}%
  \BibitemOpen
  \bibfield  {author} {\bibinfo {author} {\bibfnamefont {E.~S.~C.}\
  \bibnamefont {Ching}}, \bibinfo {author} {\bibfnamefont {J.~S.}\ \bibnamefont
  {Langer}}, \ and\ \bibinfo {author} {\bibfnamefont {H.}~\bibnamefont
  {Nakanishi}},\ }\href {\doibase 10.1103/PhysRevLett.76.1087} {\bibfield
  {journal} {\bibinfo  {journal} {Phys. Rev. Lett.}\ }\textbf {\bibinfo
  {volume} {76}},\ \bibinfo {pages} {1087} (\bibinfo {year}
  {1996})}\BibitemShut {NoStop}%
\bibitem [{\citenamefont {Adda-Bedia}(2004)}]{Adda-Bedia2004}%
  \BibitemOpen
  \bibfield  {author} {\bibinfo {author} {\bibfnamefont {M.}~\bibnamefont
  {Adda-Bedia}},\ }\href {\doibase 10.1103/PhysRevLett.93.185502} {\bibfield
  {journal} {\bibinfo  {journal} {Phys. Rev. Lett.}\ }\textbf {\bibinfo
  {volume} {93}},\ \bibinfo {pages} {185502} (\bibinfo {year}
  {2004})}\BibitemShut {NoStop}%
\bibitem [{\citenamefont {Adda-Bedia}\ \emph {et~al.}(2013)\citenamefont
  {Adda-Bedia}, \citenamefont {Arias}, \citenamefont {Bouchbinder},\ and\
  \citenamefont {Katzav}}]{Adda-Bedia2013}%
  \BibitemOpen
  \bibfield  {author} {\bibinfo {author} {\bibfnamefont {M.}~\bibnamefont
  {Adda-Bedia}}, \bibinfo {author} {\bibfnamefont {R.~E.}\ \bibnamefont
  {Arias}}, \bibinfo {author} {\bibfnamefont {E.}~\bibnamefont {Bouchbinder}},
  \ and\ \bibinfo {author} {\bibfnamefont {E.}~\bibnamefont {Katzav}},\ }\href
  {\doibase 10.1103/PhysRevLett.110.014302} {\bibfield  {journal} {\bibinfo
  {journal} {Phys. Rev. Lett.}\ }\textbf {\bibinfo {volume} {110}},\ \bibinfo
  {pages} {014302} (\bibinfo {year} {2013})}\BibitemShut {NoStop}%
\bibitem [{\citenamefont {Brener}\ and\ \citenamefont
  {Marchenko}(1998)}]{Brener1998}%
  \BibitemOpen
  \bibfield  {author} {\bibinfo {author} {\bibfnamefont {E.~A.}\ \bibnamefont
  {Brener}}\ and\ \bibinfo {author} {\bibfnamefont {V.~I.}\ \bibnamefont
  {Marchenko}},\ }\href {\doibase 10.1103/PhysRevLett.81.5141} {\bibfield
  {journal} {\bibinfo  {journal} {Phys. Rev. Lett.}\ }\textbf {\bibinfo
  {volume} {81}},\ \bibinfo {pages} {5141} (\bibinfo {year}
  {1998})}\BibitemShut {NoStop}%
\bibitem [{\citenamefont {Lund}(1996)}]{Lund1996}%
  \BibitemOpen
  \bibfield  {author} {\bibinfo {author} {\bibfnamefont {F.}~\bibnamefont
  {Lund}},\ }\href {\doibase 10.1103/PhysRevLett.76.2742} {\bibfield  {journal}
  {\bibinfo  {journal} {Phys. Rev. Lett.}\ }\textbf {\bibinfo {volume} {76}},\
  \bibinfo {pages} {2742} (\bibinfo {year} {1996})}\BibitemShut {NoStop}%
\bibitem [{\citenamefont {Bouchbinder}\ and\ \citenamefont
  {Procaccia}(2007)}]{Bouchbinder2007}%
  \BibitemOpen
  \bibfield  {author} {\bibinfo {author} {\bibfnamefont {E.}~\bibnamefont
  {Bouchbinder}}\ and\ \bibinfo {author} {\bibfnamefont {I.}~\bibnamefont
  {Procaccia}},\ }\href {\doibase 10.1103/PhysRevLett.98.124302} {\bibfield
  {journal} {\bibinfo  {journal} {Phys. Rev. Lett.}\ }\textbf {\bibinfo
  {volume} {98}},\ \bibinfo {pages} {124302} (\bibinfo {year}
  {2007})}\BibitemShut {NoStop}%
\bibitem [{\citenamefont {Deegan}\ \emph {et~al.}(2001)\citenamefont {Deegan},
  \citenamefont {Petersan}, \citenamefont {Marder},\ and\ \citenamefont
  {Swinney}}]{Deegan2002}%
  \BibitemOpen
  \bibfield  {author} {\bibinfo {author} {\bibfnamefont {R.~D.}\ \bibnamefont
  {Deegan}}, \bibinfo {author} {\bibfnamefont {P.~J.}\ \bibnamefont
  {Petersan}}, \bibinfo {author} {\bibfnamefont {M.}~\bibnamefont {Marder}}, \
  and\ \bibinfo {author} {\bibfnamefont {H.~L.}\ \bibnamefont {Swinney}},\
  }\href {\doibase 10.1103/PhysRevLett.88.014304} {\bibfield  {journal}
  {\bibinfo  {journal} {Phys. Rev. Lett.}\ }\textbf {\bibinfo {volume} {88}},\
  \bibinfo {pages} {014304} (\bibinfo {year} {2001})}\BibitemShut {NoStop}%
\bibitem [{\citenamefont {Petersan}\ \emph {et~al.}(2004)\citenamefont
  {Petersan}, \citenamefont {Deegan}, \citenamefont {Marder},\ and\
  \citenamefont {Swinney}}]{Petersan2004}%
  \BibitemOpen
  \bibfield  {author} {\bibinfo {author} {\bibfnamefont {P.~J.}\ \bibnamefont
  {Petersan}}, \bibinfo {author} {\bibfnamefont {R.~D.}\ \bibnamefont
  {Deegan}}, \bibinfo {author} {\bibfnamefont {M.}~\bibnamefont {Marder}}, \
  and\ \bibinfo {author} {\bibfnamefont {H.~L.}\ \bibnamefont {Swinney}},\
  }\href {\doibase 10.1103/PhysRevLett.93.015504} {\bibfield  {journal}
  {\bibinfo  {journal} {Phys. Rev. Lett.}\ }\textbf {\bibinfo {volume} {93}},\
  \bibinfo {pages} {015504} (\bibinfo {year} {2004})}\BibitemShut {NoStop}%
\bibitem [{\citenamefont {Persson}(1998)}]{Persson1998}%
  \BibitemOpen
  \bibfield  {author} {\bibinfo {author} {\bibfnamefont {B.~N.~J.}\
  \bibnamefont {Persson}},\ }\href {\doibase 10.1103/PhysRevLett.81.3439}
  {\bibfield  {journal} {\bibinfo  {journal} {Phys. Rev. Lett.}\ }\textbf
  {\bibinfo {volume} {81}},\ \bibinfo {pages} {3439} (\bibinfo {year}
  {1998})}\BibitemShut {NoStop}%
\bibitem [{\citenamefont {Holian}\ \emph {et~al.}(1997)\citenamefont {Holian},
  \citenamefont {Blumenfeld},\ and\ \citenamefont {Gumbsch}}]{Holian1997}%
  \BibitemOpen
  \bibfield  {author} {\bibinfo {author} {\bibfnamefont {B.~L.}\ \bibnamefont
  {Holian}}, \bibinfo {author} {\bibfnamefont {R.}~\bibnamefont {Blumenfeld}},
  \ and\ \bibinfo {author} {\bibfnamefont {P.}~\bibnamefont {Gumbsch}},\ }\href
  {\doibase 10.1103/PhysRevLett.78.78} {\bibfield  {journal} {\bibinfo
  {journal} {Phys. Rev. Lett.}\ }\textbf {\bibinfo {volume} {78}},\ \bibinfo
  {pages} {78} (\bibinfo {year} {1997})}\BibitemShut {NoStop}%
\bibitem [{\citenamefont {Hauch}\ \emph {et~al.}(1999)\citenamefont {Hauch},
  \citenamefont {Holland}, \citenamefont {Marder},\ and\ \citenamefont
  {Swinney}}]{Hauch1999}%
  \BibitemOpen
  \bibfield  {author} {\bibinfo {author} {\bibfnamefont {J.~A.}\ \bibnamefont
  {Hauch}}, \bibinfo {author} {\bibfnamefont {D.}~\bibnamefont {Holland}},
  \bibinfo {author} {\bibfnamefont {M.}~\bibnamefont {Marder}}, \ and\ \bibinfo
  {author} {\bibfnamefont {H.~L.}\ \bibnamefont {Swinney}},\ }\href {\doibase
  10.1103/PhysRevLett.82.3823} {\bibfield  {journal} {\bibinfo  {journal}
  {Phys. Rev. Lett.}\ }\textbf {\bibinfo {volume} {82}},\ \bibinfo {pages}
  {3823} (\bibinfo {year} {1999})}\BibitemShut {NoStop}%
\bibitem [{\citenamefont {Sharon}\ and\ \citenamefont
  {Fineberg}(1999)}]{Sharon1999}%
  \BibitemOpen
  \bibfield  {author} {\bibinfo {author} {\bibfnamefont {E.}~\bibnamefont
  {Sharon}}\ and\ \bibinfo {author} {\bibfnamefont {J.}~\bibnamefont
  {Fineberg}},\ }\href {\doibase 10.1038/16891} {\bibfield  {journal} {\bibinfo
   {journal} {Nature.}\ }\textbf {\bibinfo {volume} {397}},\ \bibinfo {pages}
  {333} (\bibinfo {year} {1999})}\BibitemShut {NoStop}%
\bibitem [{\citenamefont {Stroh}(1957)}]{Stroh1957}%
  \BibitemOpen
  \bibfield  {author} {\bibinfo {author} {\bibfnamefont {A.~N.}\ \bibnamefont
  {Stroh}},\ }\href {\doibase 10.1080/00018735700101406} {\bibfield  {journal}
  {\bibinfo  {journal} {Adv. Phys.}\ }\textbf {\bibinfo {volume} {6}},\
  \bibinfo {pages} {418} (\bibinfo {year} {1957})}\BibitemShut {NoStop}%
\bibitem [{\citenamefont {Sih}(1972)}]{sih1972}%
  \BibitemOpen
  \bibfield  {author} {\bibinfo {author} {\bibfnamefont {G.~C.}\ \bibnamefont
  {Sih}},\ }\href@noop {} {\emph {\bibinfo {title} {Proceedings of an
  international conference on Dynamic Crack Propagation}}}\ (\bibinfo
  {publisher} {Springer Science \& Business Media},\ \bibinfo {year}
  {1972})\BibitemShut {NoStop}%
\bibitem [{\citenamefont {Quinn}(2019)}]{Quinn2019}%
  \BibitemOpen
  \bibfield  {author} {\bibinfo {author} {\bibfnamefont {G.~D.}\ \bibnamefont
  {Quinn}},\ }\href {\doibase 10.1111/ijag.13042} {\bibfield  {journal}
  {\bibinfo  {journal} {Int. J. Appl. Glass. Sci.}\ }\textbf {\bibinfo {volume}
  {10}},\ \bibinfo {pages} {7} (\bibinfo {year} {2019})}\BibitemShut {NoStop}%
\bibitem [{\citenamefont {Ravi-Chandar}(2004)}]{Ravi-Chandar2004}%
  \BibitemOpen
  \bibfield  {author} {\bibinfo {author} {\bibfnamefont {K.}~\bibnamefont
  {Ravi-Chandar}},\ }\href@noop {} {\emph {\bibinfo {title} {Dynamic
  fracture}}}\ (\bibinfo  {publisher} {Elsevier},\ \bibinfo {year}
  {2004})\BibitemShut {NoStop}%
\bibitem [{\citenamefont {Ravi-Chandar}\ and\ \citenamefont
  {Knauss}(1984{\natexlab{a}})}]{Ravi-Chandar1984}%
  \BibitemOpen
  \bibfield  {author} {\bibinfo {author} {\bibfnamefont {K.}~\bibnamefont
  {Ravi-Chandar}}\ and\ \bibinfo {author} {\bibfnamefont {W.~G.}\ \bibnamefont
  {Knauss}},\ }\href {\doibase 10.1007/BF01157550} {\bibfield  {journal}
  {\bibinfo  {journal} {Int. J. Fracture.}\ }\textbf {\bibinfo {volume} {26}},\
  \bibinfo {pages} {141} (\bibinfo {year} {1984}{\natexlab{a}})}\BibitemShut
  {NoStop}%
\bibitem [{\citenamefont {Freund}(1990)}]{Freund1990}%
  \BibitemOpen
  \bibfield  {author} {\bibinfo {author} {\bibfnamefont {L.~B.}\ \bibnamefont
  {Freund}},\ }\href@noop {} {\emph {\bibinfo {title} {Dynamic fracture
  mechanics}}}\ (\bibinfo  {publisher} {Cambridge Univ. Press, New York},\
  \bibinfo {year} {1990})\BibitemShut {NoStop}%
\bibitem [{\citenamefont {Goldman}\ \emph {et~al.}(2010)\citenamefont
  {Goldman}, \citenamefont {Livne},\ and\ \citenamefont
  {Fineberg}}]{Goldman2010}%
  \BibitemOpen
  \bibfield  {author} {\bibinfo {author} {\bibfnamefont {T.}~\bibnamefont
  {Goldman}}, \bibinfo {author} {\bibfnamefont {A.}~\bibnamefont {Livne}}, \
  and\ \bibinfo {author} {\bibfnamefont {J.}~\bibnamefont {Fineberg}},\ }\href
  {\doibase 10.1103/PhysRevLett.104.114301} {\bibfield  {journal} {\bibinfo
  {journal} {Phys. Rev. Lett.}\ }\textbf {\bibinfo {volume} {104}},\ \bibinfo
  {pages} {114301} (\bibinfo {year} {2010})}\BibitemShut {NoStop}%
\bibitem [{\citenamefont {Chen}\ \emph {et~al.}(2017)\citenamefont {Chen},
  \citenamefont {Bouchbinder},\ and\ \citenamefont {Karma}}]{Chen2017}%
  \BibitemOpen
  \bibfield  {author} {\bibinfo {author} {\bibfnamefont {C.~H.}\ \bibnamefont
  {Chen}}, \bibinfo {author} {\bibfnamefont {E.}~\bibnamefont {Bouchbinder}}, \
  and\ \bibinfo {author} {\bibfnamefont {A.}~\bibnamefont {Karma}},\ }\href
  {\doibase 10.1038/nphys4237} {\bibfield  {journal} {\bibinfo  {journal} {Nat.
  Phys.}\ }\textbf {\bibinfo {volume} {13}},\ \bibinfo {pages} {1186} (\bibinfo
  {year} {2017})}\BibitemShut {NoStop}%
\bibitem [{\citenamefont {Lubomirsky}\ \emph {et~al.}(2018)\citenamefont
  {Lubomirsky}, \citenamefont {Chen}, \citenamefont {Karma},\ and\
  \citenamefont {Bouchbinder}}]{Lubomirsky2018}%
  \BibitemOpen
  \bibfield  {author} {\bibinfo {author} {\bibfnamefont {Y.}~\bibnamefont
  {Lubomirsky}}, \bibinfo {author} {\bibfnamefont {C.~H.}\ \bibnamefont
  {Chen}}, \bibinfo {author} {\bibfnamefont {A.}~\bibnamefont {Karma}}, \ and\
  \bibinfo {author} {\bibfnamefont {E.}~\bibnamefont {Bouchbinder}},\ }\href
  {\doibase 10.1103/PhysRevLett.121.134301} {\bibfield  {journal} {\bibinfo
  {journal} {Phys. Rev. Lett.}\ }\textbf {\bibinfo {volume} {121}},\ \bibinfo
  {pages} {134301} (\bibinfo {year} {2018})}\BibitemShut {NoStop}%
\bibitem [{\citenamefont {Livne}\ \emph {et~al.}(2005)\citenamefont {Livne},
  \citenamefont {Cohen},\ and\ \citenamefont {Fineberg}}]{Livne2005}%
  \BibitemOpen
  \bibfield  {author} {\bibinfo {author} {\bibfnamefont {A.}~\bibnamefont
  {Livne}}, \bibinfo {author} {\bibfnamefont {G.}~\bibnamefont {Cohen}}, \ and\
  \bibinfo {author} {\bibfnamefont {J.}~\bibnamefont {Fineberg}},\ }\href
  {\doibase 10.1103/PhysRevLett.94.224301} {\bibfield  {journal} {\bibinfo
  {journal} {Phys. Rev. Lett.}\ }\textbf {\bibinfo {volume} {94}},\ \bibinfo
  {pages} {224301} (\bibinfo {year} {2005})}\BibitemShut {NoStop}%
\bibitem [{\citenamefont {Gross}\ \emph {et~al.}(1993)\citenamefont {Gross},
  \citenamefont {Fineberg}, \citenamefont {Marder}, \citenamefont {McCormick},\
  and\ \citenamefont {Swinney}}]{Gross1993}%
  \BibitemOpen
  \bibfield  {author} {\bibinfo {author} {\bibfnamefont {S.~P.}\ \bibnamefont
  {Gross}}, \bibinfo {author} {\bibfnamefont {J.}~\bibnamefont {Fineberg}},
  \bibinfo {author} {\bibfnamefont {M.}~\bibnamefont {Marder}}, \bibinfo
  {author} {\bibfnamefont {W.~D.}\ \bibnamefont {McCormick}}, \ and\ \bibinfo
  {author} {\bibfnamefont {H.~L.}\ \bibnamefont {Swinney}},\ }\href {\doibase
  10.1103/PhysRevLett.71.3162} {\bibfield  {journal} {\bibinfo  {journal}
  {Phys. Rev. Lett.}\ }\textbf {\bibinfo {volume} {71}},\ \bibinfo {pages}
  {3162} (\bibinfo {year} {1993})}\BibitemShut {NoStop}%
\bibitem [{\citenamefont {Boudet}\ and\ \citenamefont
  {Ciliberto}(1998)}]{Boudet1998}%
  \BibitemOpen
  \bibfield  {author} {\bibinfo {author} {\bibfnamefont {J.~F.}\ \bibnamefont
  {Boudet}}\ and\ \bibinfo {author} {\bibfnamefont {S.}~\bibnamefont
  {Ciliberto}},\ }\href {\doibase 10.1103/PhysRevLett.80.341} {\bibfield
  {journal} {\bibinfo  {journal} {Phys. Rev. Lett.}\ }\textbf {\bibinfo
  {volume} {80}},\ \bibinfo {pages} {341} (\bibinfo {year} {1998})}\BibitemShut
  {NoStop}%
\bibitem [{\citenamefont {Sharon}\ \emph {et~al.}(2001)\citenamefont {Sharon},
  \citenamefont {Cohen},\ and\ \citenamefont {Fineberg}}]{Sharon2001}%
  \BibitemOpen
  \bibfield  {author} {\bibinfo {author} {\bibfnamefont {E.}~\bibnamefont
  {Sharon}}, \bibinfo {author} {\bibfnamefont {G.}~\bibnamefont {Cohen}}, \
  and\ \bibinfo {author} {\bibfnamefont {J.}~\bibnamefont {Fineberg}},\ }\href
  {\doibase 10.1038/35065051} {\bibfield  {journal} {\bibinfo  {journal}
  {Nature.}\ }\textbf {\bibinfo {volume} {410}},\ \bibinfo {pages} {68}
  (\bibinfo {year} {2001})}\BibitemShut {NoStop}%
\bibitem [{\citenamefont {Sharon}\ \emph {et~al.}(2002)\citenamefont {Sharon},
  \citenamefont {Cohen},\ and\ \citenamefont {Fineberg}}]{Sharon2002}%
  \BibitemOpen
  \bibfield  {author} {\bibinfo {author} {\bibfnamefont {E.}~\bibnamefont
  {Sharon}}, \bibinfo {author} {\bibfnamefont {G.}~\bibnamefont {Cohen}}, \
  and\ \bibinfo {author} {\bibfnamefont {J.}~\bibnamefont {Fineberg}},\ }\href
  {\doibase 10.1103/PhysRevLett.88.085503} {\bibfield  {journal} {\bibinfo
  {journal} {Phys. Rev. Lett.}\ }\textbf {\bibinfo {volume} {88}},\ \bibinfo
  {pages} {085503} (\bibinfo {year} {2002})}\BibitemShut {NoStop}%
\bibitem [{\citenamefont {Massy}\ \emph {et~al.}(2018)\citenamefont {Massy},
  \citenamefont {Mazen}, \citenamefont {Landru}, \citenamefont {Ben~Mohamed},
  \citenamefont {Tardif}, \citenamefont {Reinhardt}, \citenamefont {Madeira},
  \citenamefont {Kononchuk},\ and\ \citenamefont {Rieutord}}]{Massy2018}%
  \BibitemOpen
  \bibfield  {author} {\bibinfo {author} {\bibfnamefont {D.}~\bibnamefont
  {Massy}}, \bibinfo {author} {\bibfnamefont {F.}~\bibnamefont {Mazen}},
  \bibinfo {author} {\bibfnamefont {D.}~\bibnamefont {Landru}}, \bibinfo
  {author} {\bibfnamefont {N.}~\bibnamefont {Ben~Mohamed}}, \bibinfo {author}
  {\bibfnamefont {S.}~\bibnamefont {Tardif}}, \bibinfo {author} {\bibfnamefont
  {A.}~\bibnamefont {Reinhardt}}, \bibinfo {author} {\bibfnamefont
  {F.}~\bibnamefont {Madeira}}, \bibinfo {author} {\bibfnamefont
  {O.}~\bibnamefont {Kononchuk}}, \ and\ \bibinfo {author} {\bibfnamefont
  {F.}~\bibnamefont {Rieutord}},\ }\href {\doibase
  10.1103/PhysRevLett.121.195501} {\bibfield  {journal} {\bibinfo  {journal}
  {Phys. Rev. Lett.}\ }\textbf {\bibinfo {volume} {121}},\ \bibinfo {pages}
  {195501} (\bibinfo {year} {2018})}\BibitemShut {NoStop}%
\bibitem [{\citenamefont {Sharon}\ \emph {et~al.}(1996)\citenamefont {Sharon},
  \citenamefont {Gross},\ and\ \citenamefont {Fineberg}}]{Sharon1996}%
  \BibitemOpen
  \bibfield  {author} {\bibinfo {author} {\bibfnamefont {E.}~\bibnamefont
  {Sharon}}, \bibinfo {author} {\bibfnamefont {S.~P.}\ \bibnamefont {Gross}}, \
  and\ \bibinfo {author} {\bibfnamefont {J.}~\bibnamefont {Fineberg}},\ }\href
  {\doibase 10.1103/PhysRevLett.76.2117} {\bibfield  {journal} {\bibinfo
  {journal} {Phys. Rev. Lett.}\ }\textbf {\bibinfo {volume} {76}},\ \bibinfo
  {pages} {2117} (\bibinfo {year} {1996})}\BibitemShut {NoStop}%
\bibitem [{\citenamefont {Sharon}\ \emph {et~al.}(1995)\citenamefont {Sharon},
  \citenamefont {Gross},\ and\ \citenamefont {Fineberg}}]{Sharon1995}%
  \BibitemOpen
  \bibfield  {author} {\bibinfo {author} {\bibfnamefont {E.}~\bibnamefont
  {Sharon}}, \bibinfo {author} {\bibfnamefont {S.~P.}\ \bibnamefont {Gross}}, \
  and\ \bibinfo {author} {\bibfnamefont {J.}~\bibnamefont {Fineberg}},\ }\href
  {\doibase 10.1103/PhysRevLett.74.5096} {\bibfield  {journal} {\bibinfo
  {journal} {Phys. Rev. Lett.}\ }\textbf {\bibinfo {volume} {74}},\ \bibinfo
  {pages} {5096} (\bibinfo {year} {1995})}\BibitemShut {NoStop}%
\bibitem [{\citenamefont {Kobayashi}\ \emph {et~al.}(1974)\citenamefont
  {Kobayashi}, \citenamefont {Wade}, \citenamefont {Bradley},\ and\
  \citenamefont {Chiu}}]{Kobayashi1972}%
  \BibitemOpen
  \bibfield  {author} {\bibinfo {author} {\bibfnamefont {A.~S.}\ \bibnamefont
  {Kobayashi}}, \bibinfo {author} {\bibfnamefont {B.~G.}\ \bibnamefont {Wade}},
  \bibinfo {author} {\bibfnamefont {W.~B.}\ \bibnamefont {Bradley}}, \ and\
  \bibinfo {author} {\bibfnamefont {S.~T.}\ \bibnamefont {Chiu}},\ }\href
  {\doibase 10.1016/0013-7944(74)90048-4} {\bibfield  {journal} {\bibinfo
  {journal} {Eng. Fract. Mech.}\ }\textbf {\bibinfo {volume} {6}},\ \bibinfo
  {pages} {81} (\bibinfo {year} {1974})}\BibitemShut {NoStop}%
\bibitem [{\citenamefont {Fineberg}\ \emph {et~al.}(1991)\citenamefont
  {Fineberg}, \citenamefont {Gross}, \citenamefont {Marder},\ and\
  \citenamefont {Swinney}}]{Fineberg1991}%
  \BibitemOpen
  \bibfield  {author} {\bibinfo {author} {\bibfnamefont {J.}~\bibnamefont
  {Fineberg}}, \bibinfo {author} {\bibfnamefont {S.~P.}\ \bibnamefont {Gross}},
  \bibinfo {author} {\bibfnamefont {M.}~\bibnamefont {Marder}}, \ and\ \bibinfo
  {author} {\bibfnamefont {H.~L.}\ \bibnamefont {Swinney}},\ }\href {\doibase
  10.1103/PhysRevLett.67.457} {\bibfield  {journal} {\bibinfo  {journal} {Phys.
  Rev. Lett.}\ }\textbf {\bibinfo {volume} {67}},\ \bibinfo {pages} {457}
  (\bibinfo {year} {1991})}\BibitemShut {NoStop}%
\bibitem [{\citenamefont {Fineberg}\ \emph {et~al.}(1992)\citenamefont
  {Fineberg}, \citenamefont {Gross}, \citenamefont {Marder},\ and\
  \citenamefont {Swinney}}]{Fineberg1992}%
  \BibitemOpen
  \bibfield  {author} {\bibinfo {author} {\bibfnamefont {J.}~\bibnamefont
  {Fineberg}}, \bibinfo {author} {\bibfnamefont {S.~P.}\ \bibnamefont {Gross}},
  \bibinfo {author} {\bibfnamefont {M.}~\bibnamefont {Marder}}, \ and\ \bibinfo
  {author} {\bibfnamefont {H.~L.}\ \bibnamefont {Swinney}},\ }\href {\doibase
  10.1103/PhysRevB.45.5146} {\bibfield  {journal} {\bibinfo  {journal} {Phys.
  Rev. B.}\ }\textbf {\bibinfo {volume} {45}},\ \bibinfo {pages} {5146}
  (\bibinfo {year} {1992})}\BibitemShut {NoStop}%
\bibitem [{\citenamefont {Sharon}\ and\ \citenamefont
  {Fineberg}(1996)}]{Sharon199601}%
  \BibitemOpen
  \bibfield  {author} {\bibinfo {author} {\bibfnamefont {E.}~\bibnamefont
  {Sharon}}\ and\ \bibinfo {author} {\bibfnamefont {J.}~\bibnamefont
  {Fineberg}},\ }\href {\doibase 10.1103/PhysRevB.54.7128} {\bibfield
  {journal} {\bibinfo  {journal} {Phys. Rev. B}\ }\textbf {\bibinfo {volume}
  {54}},\ \bibinfo {pages} {7128} (\bibinfo {year} {1996})}\BibitemShut
  {NoStop}%
\bibitem [{\citenamefont {Hull}(1999)}]{Hull}%
  \BibitemOpen
  \bibfield  {author} {\bibinfo {author} {\bibfnamefont {D.}~\bibnamefont
  {Hull}},\ }\href@noop {} {\emph {\bibinfo {title} {Fractography: Observing,
  Measuring and interpreting fracture surface topography}}}\ (\bibinfo
  {publisher} {Cambridge Univ. Press, New York},\ \bibinfo {year}
  {1999})\BibitemShut {NoStop}%
\bibitem [{\citenamefont {Ravi-Chandar}\ and\ \citenamefont
  {Knauss}(1984{\natexlab{b}})}]{Ravi-Chandar1984II}%
  \BibitemOpen
  \bibfield  {author} {\bibinfo {author} {\bibfnamefont {K.}~\bibnamefont
  {Ravi-Chandar}}\ and\ \bibinfo {author} {\bibfnamefont {W.~G.}\ \bibnamefont
  {Knauss}},\ }\href {\doibase 10.1007/bf01152313} {\bibfield  {journal}
  {\bibinfo  {journal} {Inter. J. Fracture}\ }\textbf {\bibinfo {volume}
  {26}},\ \bibinfo {pages} {65} (\bibinfo {year}
  {1984}{\natexlab{b}})}\BibitemShut {NoStop}%
\bibitem [{\citenamefont {Ravi-Chandar}\ and\ \citenamefont
  {Yang}(1997)}]{Ravi1997}%
  \BibitemOpen
  \bibfield  {author} {\bibinfo {author} {\bibfnamefont {K.}~\bibnamefont
  {Ravi-Chandar}}\ and\ \bibinfo {author} {\bibfnamefont {B.}~\bibnamefont
  {Yang}},\ }\href {\doibase 10.1016/s0022-5096(96)00096-8} {\bibfield
  {journal} {\bibinfo  {journal} {J. Mech. Phys. Solids.}\ }\textbf {\bibinfo
  {volume} {45}},\ \bibinfo {pages} {535} (\bibinfo {year} {1997})}\BibitemShut
  {NoStop}%
\bibitem [{\citenamefont {Cramer}\ \emph {et~al.}(2000)\citenamefont {Cramer},
  \citenamefont {Wanner},\ and\ \citenamefont {Gumbsch}}]{Cramer2000}%
  \BibitemOpen
  \bibfield  {author} {\bibinfo {author} {\bibfnamefont {T.}~\bibnamefont
  {Cramer}}, \bibinfo {author} {\bibfnamefont {A.}~\bibnamefont {Wanner}}, \
  and\ \bibinfo {author} {\bibfnamefont {P.}~\bibnamefont {Gumbsch}},\ }\href
  {\doibase 10.1103/PhysRevLett.85.788} {\bibfield  {journal} {\bibinfo
  {journal} {Phys. Rev. Lett.}\ }\textbf {\bibinfo {volume} {85}},\ \bibinfo
  {pages} {788} (\bibinfo {year} {2000})}\BibitemShut {NoStop}%
\bibitem [{\citenamefont {Scheibert}\ \emph {et~al.}(2010)\citenamefont
  {Scheibert}, \citenamefont {Guerra}, \citenamefont {C\'elari\'e},
  \citenamefont {Dalmas},\ and\ \citenamefont {Bonamy}}]{Scheibert2010}%
  \BibitemOpen
  \bibfield  {author} {\bibinfo {author} {\bibfnamefont {J.}~\bibnamefont
  {Scheibert}}, \bibinfo {author} {\bibfnamefont {C.}~\bibnamefont {Guerra}},
  \bibinfo {author} {\bibfnamefont {F.}~\bibnamefont {C\'elari\'e}}, \bibinfo
  {author} {\bibfnamefont {D.}~\bibnamefont {Dalmas}}, \ and\ \bibinfo {author}
  {\bibfnamefont {D.}~\bibnamefont {Bonamy}},\ }\href {\doibase
  10.1103/PhysRevLett.104.045501} {\bibfield  {journal} {\bibinfo  {journal}
  {Phys. Rev. Lett.}\ }\textbf {\bibinfo {volume} {104}},\ \bibinfo {pages}
  {045501} (\bibinfo {year} {2010})}\BibitemShut {NoStop}%
\bibitem [{\citenamefont {Claudia}\ \emph {et~al.}(2012)\citenamefont
  {Claudia}, \citenamefont {Julien}, \citenamefont {Daniel},\ and\
  \citenamefont {Davy}}]{Claudia2012}%
  \BibitemOpen
  \bibfield  {author} {\bibinfo {author} {\bibfnamefont {G.}~\bibnamefont
  {Claudia}}, \bibinfo {author} {\bibfnamefont {S.}~\bibnamefont {Julien}},
  \bibinfo {author} {\bibfnamefont {B.}~\bibnamefont {Daniel}}, \ and\ \bibinfo
  {author} {\bibfnamefont {D.}~\bibnamefont {Davy}},\ }\href {\doibase
  10.1073/pnas.1113205109} {\bibfield  {journal} {\bibinfo  {journal} {Proc.
  Natl. Acad. Sci. USA.}\ }\textbf {\bibinfo {volume} {109}},\ \bibinfo {pages}
  {390} (\bibinfo {year} {2012})}\BibitemShut {NoStop}%
\bibitem [{\citenamefont {Livne}\ \emph {et~al.}(2007)\citenamefont {Livne},
  \citenamefont {Ben-David},\ and\ \citenamefont {Fineberg}}]{Livne2007}%
  \BibitemOpen
  \bibfield  {author} {\bibinfo {author} {\bibfnamefont {A.}~\bibnamefont
  {Livne}}, \bibinfo {author} {\bibfnamefont {O.}~\bibnamefont {Ben-David}}, \
  and\ \bibinfo {author} {\bibfnamefont {J.}~\bibnamefont {Fineberg}},\ }\href
  {\doibase 10.1103/PhysRevLett.98.124301} {\bibfield  {journal} {\bibinfo
  {journal} {Phys. Rev. Lett.}\ }\textbf {\bibinfo {volume} {98}},\ \bibinfo
  {pages} {124301} (\bibinfo {year} {2007})}\BibitemShut {NoStop}%
\bibitem [{\citenamefont {Goldman}\ \emph {et~al.}(2012)\citenamefont
  {Goldman}, \citenamefont {Harpaz}, \citenamefont {Bouchbinder},\ and\
  \citenamefont {Fineberg}}]{Goldman2012}%
  \BibitemOpen
  \bibfield  {author} {\bibinfo {author} {\bibfnamefont {T.}~\bibnamefont
  {Goldman}}, \bibinfo {author} {\bibfnamefont {R.}~\bibnamefont {Harpaz}},
  \bibinfo {author} {\bibfnamefont {E.}~\bibnamefont {Bouchbinder}}, \ and\
  \bibinfo {author} {\bibfnamefont {J.}~\bibnamefont {Fineberg}},\ }\href
  {\doibase 10.1103/PhysRevLett.108.104303} {\bibfield  {journal} {\bibinfo
  {journal} {Phys. Rev. Lett.}\ }\textbf {\bibinfo {volume} {108}},\ \bibinfo
  {pages} {104303} (\bibinfo {year} {2012})}\BibitemShut {NoStop}%
\bibitem [{\citenamefont {Inglis}(1913)}]{Inglis1913}%
  \BibitemOpen
  \bibfield  {author} {\bibinfo {author} {\bibfnamefont {C.~E.}\ \bibnamefont
  {Inglis}},\ }\href@noop {} {\bibfield  {journal} {\bibinfo  {journal} {Phil.
  Trans. R. Soc. Lond. A.}\ }\textbf {\bibinfo {volume} {55}},\ \bibinfo
  {pages} {219} (\bibinfo {year} {1913})}\BibitemShut {NoStop}%
\bibitem [{\citenamefont {Griffith}(1921)}]{Griffith1921}%
  \BibitemOpen
  \bibfield  {author} {\bibinfo {author} {\bibfnamefont {A.~A.}\ \bibnamefont
  {Griffith}},\ }\href {\doibase 10.1098/rsta.1921.0006} {\bibfield  {journal}
  {\bibinfo  {journal} {Phil. Trans. R. Soc. Lond. A.}\ }\textbf {\bibinfo
  {volume} {221}},\ \bibinfo {pages} {163} (\bibinfo {year}
  {1921})}\BibitemShut {NoStop}%
\bibitem [{\citenamefont {Fineberg}\ and\ \citenamefont
  {Bouchbinder}(2015)}]{Fineberg2015}%
  \BibitemOpen
  \bibfield  {author} {\bibinfo {author} {\bibfnamefont {J.}~\bibnamefont
  {Fineberg}}\ and\ \bibinfo {author} {\bibfnamefont {E.}~\bibnamefont
  {Bouchbinder}},\ }\href {\doibase 10.1007/s10704-015-0038-x} {\bibfield
  {journal} {\bibinfo  {journal} {Int. J. Fracture.}\ }\textbf {\bibinfo
  {volume} {196}},\ \bibinfo {pages} {33} (\bibinfo {year} {2015})}\BibitemShut
  {NoStop}%
\bibitem [{\citenamefont {Bouchbinder}\ \emph {et~al.}(2014)\citenamefont
  {Bouchbinder}, \citenamefont {Goldman},\ and\ \citenamefont
  {Fineberg}}]{Bouchbinder2014}%
  \BibitemOpen
  \bibfield  {author} {\bibinfo {author} {\bibfnamefont {E.}~\bibnamefont
  {Bouchbinder}}, \bibinfo {author} {\bibfnamefont {T.}~\bibnamefont
  {Goldman}}, \ and\ \bibinfo {author} {\bibfnamefont {J.}~\bibnamefont
  {Fineberg}},\ }\href {\doibase 10.1088/0034-4885/77/4/046501} {\bibfield
  {journal} {\bibinfo  {journal} {Rep. Prog. Phys.}\ }\textbf {\bibinfo
  {volume} {77}},\ \bibinfo {pages} {046501} (\bibinfo {year}
  {2014})}\BibitemShut {NoStop}%
\bibitem [{\citenamefont {Bouchbinder}\ \emph {et~al.}(2010)\citenamefont
  {Bouchbinder}, \citenamefont {Fineberg},\ and\ \citenamefont
  {Marder}}]{Bouchbinder2010}%
  \BibitemOpen
  \bibfield  {author} {\bibinfo {author} {\bibfnamefont {E.}~\bibnamefont
  {Bouchbinder}}, \bibinfo {author} {\bibfnamefont {J.}~\bibnamefont
  {Fineberg}}, \ and\ \bibinfo {author} {\bibfnamefont {M.}~\bibnamefont
  {Marder}},\ }\href {\doibase 10.1146/annurev-conmatphys-070909-104019}
  {\bibfield  {journal} {\bibinfo  {journal} {Annu. Rev. Condens. Matter.
  Phys.}\ }\textbf {\bibinfo {volume} {1}},\ \bibinfo {pages} {371} (\bibinfo
  {year} {2010})}\BibitemShut {NoStop}%
\bibitem [{\citenamefont {Yoffe}(1951)}]{Yoffe1951}%
  \BibitemOpen
  \bibfield  {author} {\bibinfo {author} {\bibfnamefont {E.~H.}\ \bibnamefont
  {Yoffe}},\ }\href {\doibase 10.1080/14786445108561302} {\bibfield  {journal}
  {\bibinfo  {journal} {Philos. Mag.}\ }\textbf {\bibinfo {volume} {42}},\
  \bibinfo {pages} {739} (\bibinfo {year} {1951})}\BibitemShut {NoStop}%
\bibitem [{\citenamefont {Bonamy}\ and\ \citenamefont
  {Ravi-Chandar}(2003)}]{Bonamy2003}%
  \BibitemOpen
  \bibfield  {author} {\bibinfo {author} {\bibfnamefont {D.}~\bibnamefont
  {Bonamy}}\ and\ \bibinfo {author} {\bibfnamefont {K.}~\bibnamefont
  {Ravi-Chandar}},\ }\href {\doibase 10.1103/PhysRevLett.91.235502} {\bibfield
  {journal} {\bibinfo  {journal} {Phys. Rev. Lett.}\ }\textbf {\bibinfo
  {volume} {91}},\ \bibinfo {pages} {235502} (\bibinfo {year}
  {2003})}\BibitemShut {NoStop}%
\bibitem [{\citenamefont {Bouchaud}\ \emph {et~al.}(2002)\citenamefont
  {Bouchaud}, \citenamefont {Bouchaud}, \citenamefont {Fisher}, \citenamefont
  {Ramanathan},\ and\ \citenamefont {Rice}}]{Bouchaud2002}%
  \BibitemOpen
  \bibfield  {author} {\bibinfo {author} {\bibfnamefont {E.}~\bibnamefont
  {Bouchaud}}, \bibinfo {author} {\bibfnamefont {J.~P.}\ \bibnamefont
  {Bouchaud}}, \bibinfo {author} {\bibfnamefont {D.~S.}\ \bibnamefont
  {Fisher}}, \bibinfo {author} {\bibfnamefont {S.}~\bibnamefont {Ramanathan}},
  \ and\ \bibinfo {author} {\bibfnamefont {J.~R.}\ \bibnamefont {Rice}},\
  }\href {\doibase 10.1016/S0022-5096(01)00137-5} {\bibfield  {journal}
  {\bibinfo  {journal} {J. Mech. Phys. Solids.}\ }\textbf {\bibinfo {volume}
  {50}},\ \bibinfo {pages} {1703} (\bibinfo {year} {2002})}\BibitemShut
  {NoStop}%
\bibitem [{\citenamefont {Pereira}\ \emph {et~al.}(2017)\citenamefont
  {Pereira}, \citenamefont {Weerheijm},\ and\ \citenamefont
  {Sluys}}]{Pereira2017}%
  \BibitemOpen
  \bibfield  {author} {\bibinfo {author} {\bibfnamefont {L.~F.}\ \bibnamefont
  {Pereira}}, \bibinfo {author} {\bibfnamefont {J.}~\bibnamefont {Weerheijm}},
  \ and\ \bibinfo {author} {\bibfnamefont {L.~J.}\ \bibnamefont {Sluys}},\
  }\href {\doibase 10.1016/j.engfracmech.2017.06.019} {\bibfield  {journal}
  {\bibinfo  {journal} {Eng. Fract. Mech.}\ }\textbf {\bibinfo {volume}
  {182}},\ \bibinfo {pages} {689} (\bibinfo {year} {2017})}\BibitemShut
  {NoStop}%
\bibitem [{\citenamefont {Bobaru}\ and\ \citenamefont
  {Zhang}(2015)}]{Bobaru2015}%
  \BibitemOpen
  \bibfield  {author} {\bibinfo {author} {\bibfnamefont {F.}~\bibnamefont
  {Bobaru}}\ and\ \bibinfo {author} {\bibfnamefont {G.}~\bibnamefont {Zhang}},\
  }\href {\doibase 10.1007/s10704-015-0056-8} {\bibfield  {journal} {\bibinfo
  {journal} {Int. J. Fracture.}\ }\textbf {\bibinfo {volume} {196}},\ \bibinfo
  {pages} {59} (\bibinfo {year} {2015})}\BibitemShut {NoStop}%
\bibitem [{\citenamefont {Livne}\ \emph {et~al.}(2008)\citenamefont {Livne},
  \citenamefont {Bouchbinder},\ and\ \citenamefont {Fineberg}}]{Livne2008}%
  \BibitemOpen
  \bibfield  {author} {\bibinfo {author} {\bibfnamefont {A.}~\bibnamefont
  {Livne}}, \bibinfo {author} {\bibfnamefont {E.}~\bibnamefont {Bouchbinder}},
  \ and\ \bibinfo {author} {\bibfnamefont {J.}~\bibnamefont {Fineberg}},\
  }\href {\doibase 10.1103/PhysRevLett.101.264301} {\bibfield  {journal}
  {\bibinfo  {journal} {Phys. Rev. Lett.}\ }\textbf {\bibinfo {volume} {101}},\
  \bibinfo {pages} {264301} (\bibinfo {year} {2008})}\BibitemShut {NoStop}%
\bibitem [{\citenamefont {Bouchbinder}\ \emph {et~al.}(2008)\citenamefont
  {Bouchbinder}, \citenamefont {Livne},\ and\ \citenamefont
  {Fineberg}}]{Bouchbinder2008}%
  \BibitemOpen
  \bibfield  {author} {\bibinfo {author} {\bibfnamefont {E.}~\bibnamefont
  {Bouchbinder}}, \bibinfo {author} {\bibfnamefont {A.}~\bibnamefont {Livne}},
  \ and\ \bibinfo {author} {\bibfnamefont {J.}~\bibnamefont {Fineberg}},\
  }\href {\doibase 10.1103/PhysRevLett.101.264302} {\bibfield  {journal}
  {\bibinfo  {journal} {Phys. Rev. Lett.}\ }\textbf {\bibinfo {volume} {101}},\
  \bibinfo {pages} {264302} (\bibinfo {year} {2008})}\BibitemShut {NoStop}%
\bibitem [{\citenamefont {Bouchbinder}(2009)}]{Bouchbinder2009}%
  \BibitemOpen
  \bibfield  {author} {\bibinfo {author} {\bibfnamefont {E.}~\bibnamefont
  {Bouchbinder}},\ }\href {\doibase 10.1103/PhysRevLett.103.164301} {\bibfield
  {journal} {\bibinfo  {journal} {Phys. Rev. Lett.}\ }\textbf {\bibinfo
  {volume} {103}},\ \bibinfo {pages} {164301} (\bibinfo {year}
  {2009})}\BibitemShut {NoStop}%
\bibitem [{\citenamefont {Livne}\ \emph {et~al.}(2010)\citenamefont {Livne},
  \citenamefont {Bouchbinder}, \citenamefont {Svetlizky},\ and\ \citenamefont
  {Fineberg}}]{livne2010}%
  \BibitemOpen
  \bibfield  {author} {\bibinfo {author} {\bibfnamefont {A.}~\bibnamefont
  {Livne}}, \bibinfo {author} {\bibfnamefont {E.}~\bibnamefont {Bouchbinder}},
  \bibinfo {author} {\bibfnamefont {I.}~\bibnamefont {Svetlizky}}, \ and\
  \bibinfo {author} {\bibfnamefont {J.}~\bibnamefont {Fineberg}},\ }\href
  {\doibase 10.1126/science.1180476} {\bibfield  {journal} {\bibinfo  {journal}
  {Science.}\ }\textbf {\bibinfo {volume} {327}},\ \bibinfo {pages} {1359}
  (\bibinfo {year} {2010})}\BibitemShut {NoStop}%
\bibitem [{\citenamefont {Buehler}\ \emph {et~al.}(2003)\citenamefont
  {Buehler}, \citenamefont {Abraham},\ and\ \citenamefont {Gao}}]{Buehler2003}%
  \BibitemOpen
  \bibfield  {author} {\bibinfo {author} {\bibfnamefont {M.~J.}\ \bibnamefont
  {Buehler}}, \bibinfo {author} {\bibfnamefont {F.~F.}\ \bibnamefont
  {Abraham}}, \ and\ \bibinfo {author} {\bibfnamefont {H.}~\bibnamefont
  {Gao}},\ }\href {\doibase 10.1038/nature02096} {\bibfield  {journal}
  {\bibinfo  {journal} {Nature.}\ }\textbf {\bibinfo {volume} {426}},\ \bibinfo
  {pages} {141} (\bibinfo {year} {2003})}\BibitemShut {NoStop}%
\bibitem [{\citenamefont {Buehler}\ and\ \citenamefont
  {Gao}(2006)}]{Buehler2006}%
  \BibitemOpen
  \bibfield  {author} {\bibinfo {author} {\bibfnamefont {M.~J.}\ \bibnamefont
  {Buehler}}\ and\ \bibinfo {author} {\bibfnamefont {H.}~\bibnamefont {Gao}},\
  }\href {\doibase 10.1038/nature04408} {\bibfield  {journal} {\bibinfo
  {journal} {Nature.}\ }\textbf {\bibinfo {volume} {929}},\ \bibinfo {pages}
  {307} (\bibinfo {year} {2006})}\BibitemShut {NoStop}%
\bibitem [{\citenamefont {Abraham}(1996)}]{Abraham1996}%
  \BibitemOpen
  \bibfield  {author} {\bibinfo {author} {\bibfnamefont {F.~F.}\ \bibnamefont
  {Abraham}},\ }\href {\doibase 10.1103/PhysRevLett.77.869} {\bibfield
  {journal} {\bibinfo  {journal} {Phys. Rev. Lett.}\ }\textbf {\bibinfo
  {volume} {77}},\ \bibinfo {pages} {869} (\bibinfo {year} {1996})}\BibitemShut
  {NoStop}%
\bibitem [{\citenamefont {Buehler}\ \emph {et~al.}(2007)\citenamefont
  {Buehler}, \citenamefont {Tang}, \citenamefont {van Duin},\ and\
  \citenamefont {Goddard}}]{Buehler2007}%
  \BibitemOpen
  \bibfield  {author} {\bibinfo {author} {\bibfnamefont {M.~J.}\ \bibnamefont
  {Buehler}}, \bibinfo {author} {\bibfnamefont {H.}~\bibnamefont {Tang}},
  \bibinfo {author} {\bibfnamefont {A.~C.~T.}\ \bibnamefont {van Duin}}, \ and\
  \bibinfo {author} {\bibfnamefont {W.~A.}\ \bibnamefont {Goddard}},\ }\href
  {\doibase 10.1103/PhysRevLett.99.165502} {\bibfield  {journal} {\bibinfo
  {journal} {Phys. Rev. Lett.}\ }\textbf {\bibinfo {volume} {99}},\ \bibinfo
  {pages} {165502} (\bibinfo {year} {2007})}\BibitemShut {NoStop}%
\bibitem [{\citenamefont {Boulbitch}\ and\ \citenamefont
  {Korzhenevskii}(2011)}]{Boulbitch2011}%
  \BibitemOpen
  \bibfield  {author} {\bibinfo {author} {\bibfnamefont {A.}~\bibnamefont
  {Boulbitch}}\ and\ \bibinfo {author} {\bibfnamefont {A.~L.}\ \bibnamefont
  {Korzhenevskii}},\ }\href {\doibase 10.1103/PhysRevLett.107.085505}
  {\bibfield  {journal} {\bibinfo  {journal} {Phys. Rev. Lett.}\ }\textbf
  {\bibinfo {volume} {107}},\ \bibinfo {pages} {085505} (\bibinfo {year}
  {2011})}\BibitemShut {NoStop}%
\bibitem [{\citenamefont {Goldman}\ \emph {et~al.}(2015)\citenamefont
  {Goldman}, \citenamefont {Cohen},\ and\ \citenamefont
  {Fineberg}}]{Goldman2015}%
  \BibitemOpen
  \bibfield  {author} {\bibinfo {author} {\bibfnamefont {T.}~\bibnamefont
  {Goldman}}, \bibinfo {author} {\bibfnamefont {G.}~\bibnamefont {Cohen}}, \
  and\ \bibinfo {author} {\bibfnamefont {J.}~\bibnamefont {Fineberg}},\ }\href
  {\doibase 10.1103/PhysRevLett.114.054301} {\bibfield  {journal} {\bibinfo
  {journal} {Phys. Rev. Lett.}\ }\textbf {\bibinfo {volume} {114}},\ \bibinfo
  {pages} {054301} (\bibinfo {year} {2015})}\BibitemShut {NoStop}%
\bibitem [{\citenamefont {Adda-Bedia}\ and\ \citenamefont
  {Ben~Amar}(1996)}]{Adda-Bedia1996}%
  \BibitemOpen
  \bibfield  {author} {\bibinfo {author} {\bibfnamefont {M.}~\bibnamefont
  {Adda-Bedia}}\ and\ \bibinfo {author} {\bibfnamefont {M.}~\bibnamefont
  {Ben~Amar}},\ }\href {\doibase 10.1103/PhysRevLett.76.1497} {\bibfield
  {journal} {\bibinfo  {journal} {Phys. Rev. Lett.}\ }\textbf {\bibinfo
  {volume} {76}},\ \bibinfo {pages} {1497} (\bibinfo {year}
  {1996})}\BibitemShut {NoStop}%
\bibitem [{\citenamefont {Adda-Bedia}\ \emph {et~al.}(1999)\citenamefont
  {Adda-Bedia}, \citenamefont {Arias}, \citenamefont {Ben~Amar},\ and\
  \citenamefont {Lund}}]{Adda-Bedia1999}%
  \BibitemOpen
  \bibfield  {author} {\bibinfo {author} {\bibfnamefont {M.}~\bibnamefont
  {Adda-Bedia}}, \bibinfo {author} {\bibfnamefont {R.}~\bibnamefont {Arias}},
  \bibinfo {author} {\bibfnamefont {M.}~\bibnamefont {Ben~Amar}}, \ and\
  \bibinfo {author} {\bibfnamefont {F.}~\bibnamefont {Lund}},\ }\href {\doibase
  10.1103/PhysRevLett.82.2314} {\bibfield  {journal} {\bibinfo  {journal}
  {Phys. Rev. Lett.}\ }\textbf {\bibinfo {volume} {82}},\ \bibinfo {pages}
  {2314} (\bibinfo {year} {1999})}\BibitemShut {NoStop}%
\bibitem [{\citenamefont {Sander}\ and\ \citenamefont
  {Ghaisas}(1999)}]{Sander1999}%
  \BibitemOpen
  \bibfield  {author} {\bibinfo {author} {\bibfnamefont {L.~M.}\ \bibnamefont
  {Sander}}\ and\ \bibinfo {author} {\bibfnamefont {S.~V.}\ \bibnamefont
  {Ghaisas}},\ }\href {\doibase 10.1103/PhysRevLett.83.1994} {\bibfield
  {journal} {\bibinfo  {journal} {Phys. Rev. Lett.}\ }\textbf {\bibinfo
  {volume} {83}},\ \bibinfo {pages} {1994} (\bibinfo {year}
  {1999})}\BibitemShut {NoStop}%
\bibitem [{\citenamefont {Marder}\ and\ \citenamefont
  {Liu}(1993)}]{Marder1993}%
  \BibitemOpen
  \bibfield  {author} {\bibinfo {author} {\bibfnamefont {M.}~\bibnamefont
  {Marder}}\ and\ \bibinfo {author} {\bibfnamefont {X.}~\bibnamefont {Liu}},\
  }\href {\doibase 10.1103/PhysRevLett.71.2417} {\bibfield  {journal} {\bibinfo
   {journal} {Phys. Rev. Lett.}\ }\textbf {\bibinfo {volume} {71}},\ \bibinfo
  {pages} {2417} (\bibinfo {year} {1993})}\BibitemShut {NoStop}%
\bibitem [{\citenamefont {Marder}\ and\ \citenamefont
  {Gross}(1995)}]{Marder1995}%
  \BibitemOpen
  \bibfield  {author} {\bibinfo {author} {\bibfnamefont {M.}~\bibnamefont
  {Marder}}\ and\ \bibinfo {author} {\bibfnamefont {S.}~\bibnamefont {Gross}},\
  }\href {\doibase 10.1016/0022-5096(94)00060-I} {\bibfield  {journal}
  {\bibinfo  {journal} {J. Mech. Phys. Solids.}\ }\textbf {\bibinfo {volume}
  {43}},\ \bibinfo {pages} {1} (\bibinfo {year} {1995})}\BibitemShut {NoStop}%
\bibitem [{\citenamefont {Abraham}\ \emph {et~al.}(1994)\citenamefont
  {Abraham}, \citenamefont {Brodbeck}, \citenamefont {Rafey},\ and\
  \citenamefont {Rudge}}]{Abraham1994}%
  \BibitemOpen
  \bibfield  {author} {\bibinfo {author} {\bibfnamefont {F.~F.}\ \bibnamefont
  {Abraham}}, \bibinfo {author} {\bibfnamefont {D.}~\bibnamefont {Brodbeck}},
  \bibinfo {author} {\bibfnamefont {R.~A.}\ \bibnamefont {Rafey}}, \ and\
  \bibinfo {author} {\bibfnamefont {W.~E.}\ \bibnamefont {Rudge}},\ }\href
  {\doibase 10.1103/PhysRevLett.73.272} {\bibfield  {journal} {\bibinfo
  {journal} {Phys. Rev. Lett.}\ }\textbf {\bibinfo {volume} {73}},\ \bibinfo
  {pages} {272} (\bibinfo {year} {1994})}\BibitemShut {NoStop}%
\bibitem [{\citenamefont {Nakano}\ \emph {et~al.}(1995)\citenamefont {Nakano},
  \citenamefont {Kalia},\ and\ \citenamefont {Vashishta}}]{Nakano1995}%
  \BibitemOpen
  \bibfield  {author} {\bibinfo {author} {\bibfnamefont {A.}~\bibnamefont
  {Nakano}}, \bibinfo {author} {\bibfnamefont {R.~K.}\ \bibnamefont {Kalia}}, \
  and\ \bibinfo {author} {\bibfnamefont {P.}~\bibnamefont {Vashishta}},\ }\href
  {\doibase 10.1103/PhysRevLett.75.3138} {\bibfield  {journal} {\bibinfo
  {journal} {Phys. Rev. Lett.}\ }\textbf {\bibinfo {volume} {75}},\ \bibinfo
  {pages} {3138} (\bibinfo {year} {1995})}\BibitemShut {NoStop}%
\bibitem [{\citenamefont {Zhou}\ \emph {et~al.}(1996)\citenamefont {Zhou},
  \citenamefont {Lomdahl}, \citenamefont {Thomson},\ and\ \citenamefont
  {Holian}}]{Zhou1996}%
  \BibitemOpen
  \bibfield  {author} {\bibinfo {author} {\bibfnamefont {S.~J.}\ \bibnamefont
  {Zhou}}, \bibinfo {author} {\bibfnamefont {P.~S.}\ \bibnamefont {Lomdahl}},
  \bibinfo {author} {\bibfnamefont {R.}~\bibnamefont {Thomson}}, \ and\
  \bibinfo {author} {\bibfnamefont {B.~L.}\ \bibnamefont {Holian}},\ }\href
  {\doibase 10.1103/PhysRevLett.76.2318} {\bibfield  {journal} {\bibinfo
  {journal} {Phys. Rev. Lett.}\ }\textbf {\bibinfo {volume} {76}},\ \bibinfo
  {pages} {2318} (\bibinfo {year} {1996})}\BibitemShut {NoStop}%
\bibitem [{\citenamefont {Omeltchenko}\ \emph {et~al.}(1997)\citenamefont
  {Omeltchenko}, \citenamefont {Yu}, \citenamefont {Kalia},\ and\ \citenamefont
  {Vashishta}}]{Omeltchenko1997}%
  \BibitemOpen
  \bibfield  {author} {\bibinfo {author} {\bibfnamefont {A.}~\bibnamefont
  {Omeltchenko}}, \bibinfo {author} {\bibfnamefont {J.}~\bibnamefont {Yu}},
  \bibinfo {author} {\bibfnamefont {R.~K.}\ \bibnamefont {Kalia}}, \ and\
  \bibinfo {author} {\bibfnamefont {P.}~\bibnamefont {Vashishta}},\ }\href
  {\doibase 10.1103/PhysRevLett.78.2148} {\bibfield  {journal} {\bibinfo
  {journal} {Phys. Rev. Lett.}\ }\textbf {\bibinfo {volume} {78}},\ \bibinfo
  {pages} {2148} (\bibinfo {year} {1997})}\BibitemShut {NoStop}%
\bibitem [{\citenamefont {Abraham}\ \emph {et~al.}(1997)\citenamefont
  {Abraham}, \citenamefont {Brodbeck}, \citenamefont {Rudge},\ and\
  \citenamefont {Xu}}]{Abraham1997}%
  \BibitemOpen
  \bibfield  {author} {\bibinfo {author} {\bibfnamefont {F.~F.}\ \bibnamefont
  {Abraham}}, \bibinfo {author} {\bibfnamefont {D.}~\bibnamefont {Brodbeck}},
  \bibinfo {author} {\bibfnamefont {W.~E.}\ \bibnamefont {Rudge}}, \ and\
  \bibinfo {author} {\bibfnamefont {X.~P.}\ \bibnamefont {Xu}},\ }\href
  {\doibase 10.1016/s0022-5096(96)00103-2} {\bibfield  {journal} {\bibinfo
  {journal} {J. Mech. Phys. Solids.}\ }\textbf {\bibinfo {volume} {45}},\
  \bibinfo {pages} {1595} (\bibinfo {year} {1997})}\BibitemShut {NoStop}%
\bibitem [{\citenamefont {Holland}\ and\ \citenamefont
  {Marder}(1998)}]{Holland1998}%
  \BibitemOpen
  \bibfield  {author} {\bibinfo {author} {\bibfnamefont {D.}~\bibnamefont
  {Holland}}\ and\ \bibinfo {author} {\bibfnamefont {M.}~\bibnamefont
  {Marder}},\ }\href {\doibase 10.1103/PhysRevLett.80.746} {\bibfield
  {journal} {\bibinfo  {journal} {Phys. Rev. Lett.}\ }\textbf {\bibinfo
  {volume} {80}},\ \bibinfo {pages} {746} (\bibinfo {year} {1998})}\BibitemShut
  {NoStop}%
\bibitem [{\citenamefont {Yamakov}\ \emph {et~al.}(2005)\citenamefont
  {Yamakov}, \citenamefont {Saether}, \citenamefont {Phillips},\ and\
  \citenamefont {Glaessgen}}]{Yamakov2005}%
  \BibitemOpen
  \bibfield  {author} {\bibinfo {author} {\bibfnamefont {V.}~\bibnamefont
  {Yamakov}}, \bibinfo {author} {\bibfnamefont {E.}~\bibnamefont {Saether}},
  \bibinfo {author} {\bibfnamefont {D.~R.}\ \bibnamefont {Phillips}}, \ and\
  \bibinfo {author} {\bibfnamefont {E.~H.}\ \bibnamefont {Glaessgen}},\ }\href
  {\doibase 10.1103/PhysRevLett.95.015502} {\bibfield  {journal} {\bibinfo
  {journal} {Phys. Rev. Lett.}\ }\textbf {\bibinfo {volume} {95}},\ \bibinfo
  {pages} {015502} (\bibinfo {year} {2005})}\BibitemShut {NoStop}%
\bibitem [{\citenamefont {Swadener}\ \emph {et~al.}(2002)\citenamefont
  {Swadener}, \citenamefont {Baskes},\ and\ \citenamefont
  {Nastasi}}]{Swadener2002}%
  \BibitemOpen
  \bibfield  {author} {\bibinfo {author} {\bibfnamefont {J.~G.}\ \bibnamefont
  {Swadener}}, \bibinfo {author} {\bibfnamefont {M.~I.}\ \bibnamefont
  {Baskes}}, \ and\ \bibinfo {author} {\bibfnamefont {M.}~\bibnamefont
  {Nastasi}},\ }\href {\doibase 10.1103/PhysRevLett.89.085503} {\bibfield
  {journal} {\bibinfo  {journal} {Phys. Rev. Lett.}\ }\textbf {\bibinfo
  {volume} {89}},\ \bibinfo {pages} {085503} (\bibinfo {year}
  {2002})}\BibitemShut {NoStop}%
\bibitem [{\citenamefont {Abraham}(2003)}]{Abraham2003}%
  \BibitemOpen
  \bibfield  {author} {\bibinfo {author} {\bibfnamefont {F.~F.}\ \bibnamefont
  {Abraham}},\ }\href {\doibase 10.1080/00018730310001594198} {\bibfield
  {journal} {\bibinfo  {journal} {Adv. Phys.}\ }\textbf {\bibinfo {volume}
  {52}},\ \bibinfo {pages} {727} (\bibinfo {year} {2003})}\BibitemShut
  {NoStop}%
\bibitem [{\citenamefont {Buehler}\ \emph {et~al.}(2006)\citenamefont
  {Buehler}, \citenamefont {van Duin},\ and\ \citenamefont
  {Goddard}}]{Buehler200601}%
  \BibitemOpen
  \bibfield  {author} {\bibinfo {author} {\bibfnamefont {M.~J.}\ \bibnamefont
  {Buehler}}, \bibinfo {author} {\bibfnamefont {A.~C.~T.}\ \bibnamefont {van
  Duin}}, \ and\ \bibinfo {author} {\bibfnamefont {W.~A.}\ \bibnamefont
  {Goddard}},\ }\href {\doibase 10.1103/PhysRevLett.96.095505} {\bibfield
  {journal} {\bibinfo  {journal} {Phys. Rev. Lett.}\ }\textbf {\bibinfo
  {volume} {96}},\ \bibinfo {pages} {095505} (\bibinfo {year}
  {2006})}\BibitemShut {NoStop}%
\bibitem [{\citenamefont {Atrash}\ \emph {et~al.}(2011)\citenamefont {Atrash},
  \citenamefont {Hashibon}, \citenamefont {Gumbsch},\ and\ \citenamefont
  {Sherman}}]{Atrash2011}%
  \BibitemOpen
  \bibfield  {author} {\bibinfo {author} {\bibfnamefont {F.}~\bibnamefont
  {Atrash}}, \bibinfo {author} {\bibfnamefont {A.}~\bibnamefont {Hashibon}},
  \bibinfo {author} {\bibfnamefont {P.}~\bibnamefont {Gumbsch}}, \ and\
  \bibinfo {author} {\bibfnamefont {D.}~\bibnamefont {Sherman}},\ }\href
  {\doibase 10.1103/PhysRevLett.106.085502} {\bibfield  {journal} {\bibinfo
  {journal} {Phys. Rev. Lett.}\ }\textbf {\bibinfo {volume} {106}},\ \bibinfo
  {pages} {085502} (\bibinfo {year} {2011})}\BibitemShut {NoStop}%
\bibitem [{\citenamefont {Aranson}\ \emph {et~al.}(2000)\citenamefont
  {Aranson}, \citenamefont {Kalatsky},\ and\ \citenamefont
  {Vinokur}}]{Aranson2000}%
  \BibitemOpen
  \bibfield  {author} {\bibinfo {author} {\bibfnamefont {I.~S.}\ \bibnamefont
  {Aranson}}, \bibinfo {author} {\bibfnamefont {V.~A.}\ \bibnamefont
  {Kalatsky}}, \ and\ \bibinfo {author} {\bibfnamefont {V.~M.}\ \bibnamefont
  {Vinokur}},\ }\href {\doibase 10.1103/PhysRevLett.85.118} {\bibfield
  {journal} {\bibinfo  {journal} {Phys. Rev. Lett.}\ }\textbf {\bibinfo
  {volume} {85}},\ \bibinfo {pages} {118} (\bibinfo {year} {2000})}\BibitemShut
  {NoStop}%
\bibitem [{\citenamefont {Karma}\ and\ \citenamefont
  {Lobkovsky}(2004)}]{Karma2004}%
  \BibitemOpen
  \bibfield  {author} {\bibinfo {author} {\bibfnamefont {A.}~\bibnamefont
  {Karma}}\ and\ \bibinfo {author} {\bibfnamefont {A.~E.}\ \bibnamefont
  {Lobkovsky}},\ }\href {\doibase 10.1103/PhysRevLett.92.245510} {\bibfield
  {journal} {\bibinfo  {journal} {Phys. Rev. Lett.}\ }\textbf {\bibinfo
  {volume} {92}},\ \bibinfo {pages} {245510} (\bibinfo {year}
  {2004})}\BibitemShut {NoStop}%
\bibitem [{\citenamefont {Henry}\ and\ \citenamefont
  {Levine}(2004)}]{Henry2004}%
  \BibitemOpen
  \bibfield  {author} {\bibinfo {author} {\bibfnamefont {H.}~\bibnamefont
  {Henry}}\ and\ \bibinfo {author} {\bibfnamefont {H.}~\bibnamefont {Levine}},\
  }\href {\doibase 10.1103/PhysRevLett.93.105504} {\bibfield  {journal}
  {\bibinfo  {journal} {Phys. Rev. Lett.}\ }\textbf {\bibinfo {volume} {93}},\
  \bibinfo {pages} {105504} (\bibinfo {year} {2004})}\BibitemShut {NoStop}%
\bibitem [{\citenamefont {Bleyer}\ and\ \citenamefont
  {Molinari}(2017)}]{Bleyer2017}%
  \BibitemOpen
  \bibfield  {author} {\bibinfo {author} {\bibfnamefont {J.}~\bibnamefont
  {Bleyer}}\ and\ \bibinfo {author} {\bibfnamefont {J.~F.}\ \bibnamefont
  {Molinari}},\ }\href {\doibase 10.1063/1.4980064} {\bibfield  {journal}
  {\bibinfo  {journal} {Appl. Phys. Lett.}\ }\textbf {\bibinfo {volume}
  {110}},\ \bibinfo {pages} {333} (\bibinfo {year} {2017})}\BibitemShut
  {NoStop}%
\bibitem [{\citenamefont {Spatschek}\ \emph {et~al.}(2006)\citenamefont
  {Spatschek}, \citenamefont {Hartmann}, \citenamefont {Brener}, \citenamefont
  {M\"uller-Krumbhaar},\ and\ \citenamefont {Kassner}}]{Spatschek2006}%
  \BibitemOpen
  \bibfield  {author} {\bibinfo {author} {\bibfnamefont {R.}~\bibnamefont
  {Spatschek}}, \bibinfo {author} {\bibfnamefont {M.}~\bibnamefont {Hartmann}},
  \bibinfo {author} {\bibfnamefont {E.}~\bibnamefont {Brener}}, \bibinfo
  {author} {\bibfnamefont {H.}~\bibnamefont {M\"uller-Krumbhaar}}, \ and\
  \bibinfo {author} {\bibfnamefont {K.}~\bibnamefont {Kassner}},\ }\href
  {\doibase 10.1103/PhysRevLett.96.015502} {\bibfield  {journal} {\bibinfo
  {journal} {Phys. Rev. Lett.}\ }\textbf {\bibinfo {volume} {96}},\ \bibinfo
  {pages} {015502} (\bibinfo {year} {2006})}\BibitemShut {NoStop}%
\bibitem [{\citenamefont {Xu}\ and\ \citenamefont {Needleman}(1994)}]{Xu1994}%
  \BibitemOpen
  \bibfield  {author} {\bibinfo {author} {\bibfnamefont {X.~P.}\ \bibnamefont
  {Xu}}\ and\ \bibinfo {author} {\bibfnamefont {A.}~\bibnamefont {Needleman}},\
  }\href {\doibase 10.1016/0022-5096(94)90003-5} {\bibfield  {journal}
  {\bibinfo  {journal} {J. Mech. Phys. Solids.}\ }\textbf {\bibinfo {volume}
  {42}},\ \bibinfo {pages} {1397} (\bibinfo {year} {1994})}\BibitemShut
  {NoStop}%
\bibitem [{\citenamefont {Nguyen}\ \emph {et~al.}(2004)\citenamefont {Nguyen},
  \citenamefont {Govindjee}, \citenamefont {Klein},\ and\ \citenamefont
  {Gao}}]{Nguyen2004}%
  \BibitemOpen
  \bibfield  {author} {\bibinfo {author} {\bibfnamefont {T.~D.}\ \bibnamefont
  {Nguyen}}, \bibinfo {author} {\bibfnamefont {S.}~\bibnamefont {Govindjee}},
  \bibinfo {author} {\bibfnamefont {P.~A.}\ \bibnamefont {Klein}}, \ and\
  \bibinfo {author} {\bibfnamefont {H.}~\bibnamefont {Gao}},\ }\href {\doibase
  10.1016/j.cma.2003.09.024} {\bibfield  {journal} {\bibinfo  {journal}
  {Comput. Method. Appl. M.}\ }\textbf {\bibinfo {volume} {193}},\ \bibinfo
  {pages} {3239} (\bibinfo {year} {2004})}\BibitemShut {NoStop}%
\bibitem [{\citenamefont {Elmukashfi}\ and\ \citenamefont
  {Kroon}(2012)}]{Elmukashfi2012}%
  \BibitemOpen
  \bibfield  {author} {\bibinfo {author} {\bibfnamefont {E.}~\bibnamefont
  {Elmukashfi}}\ and\ \bibinfo {author} {\bibfnamefont {M.}~\bibnamefont
  {Kroon}},\ }\href {\doibase 10.1007/s10704-012-9761-8} {\bibfield  {journal}
  {\bibinfo  {journal} {Int. J. Fracture.}\ }\textbf {\bibinfo {volume}
  {177}},\ \bibinfo {pages} {163} (\bibinfo {year} {2012})}\BibitemShut
  {NoStop}%
\bibitem [{\citenamefont {Belytschko}\ \emph {et~al.}(2003)\citenamefont
  {Belytschko}, \citenamefont {Chen}, \citenamefont {Xu},\ and\ \citenamefont
  {Zi}}]{Belytschko2003}%
  \BibitemOpen
  \bibfield  {author} {\bibinfo {author} {\bibfnamefont {T.}~\bibnamefont
  {Belytschko}}, \bibinfo {author} {\bibfnamefont {H.}~\bibnamefont {Chen}},
  \bibinfo {author} {\bibfnamefont {J.}~\bibnamefont {Xu}}, \ and\ \bibinfo
  {author} {\bibfnamefont {G.}~\bibnamefont {Zi}},\ }\href {\doibase
  10.1002/nme.941} {\bibfield  {journal} {\bibinfo  {journal} {Int. J. Numer.
  Meth. Eng.}\ }\textbf {\bibinfo {volume} {58}},\ \bibinfo {pages} {1873}
  (\bibinfo {year} {2003})}\BibitemShut {NoStop}%
\bibitem [{\citenamefont {Hellemans}(1998)}]{Hellemans1998}%
  \BibitemOpen
  \bibfield  {author} {\bibinfo {author} {\bibfnamefont {A.}~\bibnamefont
  {Hellemans}},\ }\href {\doibase 10.1126/science.281.5379.943} {\bibfield
  {journal} {\bibinfo  {journal} {Science.}\ }\textbf {\bibinfo {volume}
  {281}},\ \bibinfo {pages} {943} (\bibinfo {year} {1998})}\BibitemShut
  {NoStop}%
\bibitem [{\citenamefont {Yuse}\ and\ \citenamefont {Sano}(1993)}]{Yuse1993}%
  \BibitemOpen
  \bibfield  {author} {\bibinfo {author} {\bibfnamefont {A.}~\bibnamefont
  {Yuse}}\ and\ \bibinfo {author} {\bibfnamefont {M.}~\bibnamefont {Sano}},\
  }\href {\doibase 10.1038/362329a0} {\bibfield  {journal} {\bibinfo  {journal}
  {Nature.}\ }\textbf {\bibinfo {volume} {362}},\ \bibinfo {pages} {329}
  (\bibinfo {year} {1993})}\BibitemShut {NoStop}%
\bibitem [{\citenamefont {Ronsin}\ \emph {et~al.}(1995)\citenamefont {Ronsin},
  \citenamefont {Heslot},\ and\ \citenamefont {Perrin}}]{Ronsin1995}%
  \BibitemOpen
  \bibfield  {author} {\bibinfo {author} {\bibfnamefont {O.}~\bibnamefont
  {Ronsin}}, \bibinfo {author} {\bibfnamefont {F.}~\bibnamefont {Heslot}}, \
  and\ \bibinfo {author} {\bibfnamefont {B.}~\bibnamefont {Perrin}},\ }\href
  {\doibase 10.1103/PhysRevLett.75.2352} {\bibfield  {journal} {\bibinfo
  {journal} {Phys. Rev. Lett.}\ }\textbf {\bibinfo {volume} {75}},\ \bibinfo
  {pages} {2352} (\bibinfo {year} {1995})}\BibitemShut {NoStop}%
\bibitem [{\citenamefont {Yuse}\ and\ \citenamefont {Sano}(1997)}]{Yuse1997}%
  \BibitemOpen
  \bibfield  {author} {\bibinfo {author} {\bibfnamefont {A.}~\bibnamefont
  {Yuse}}\ and\ \bibinfo {author} {\bibfnamefont {M.}~\bibnamefont {Sano}},\
  }\href {\doibase 10.1016/s0167-2789(97)00011-0} {\bibfield  {journal}
  {\bibinfo  {journal} {Physica. D.}\ }\textbf {\bibinfo {volume} {108}},\
  \bibinfo {pages} {365} (\bibinfo {year} {1997})}\BibitemShut {NoStop}%
\bibitem [{\citenamefont {Ronsin}\ and\ \citenamefont
  {Perrin}(1998)}]{Ronsin1998}%
  \BibitemOpen
  \bibfield  {author} {\bibinfo {author} {\bibfnamefont {O.}~\bibnamefont
  {Ronsin}}\ and\ \bibinfo {author} {\bibfnamefont {B.}~\bibnamefont
  {Perrin}},\ }\href {\doibase 10.1103/PhysRevE.58.7878} {\bibfield  {journal}
  {\bibinfo  {journal} {Phys. Rev. E.}\ }\textbf {\bibinfo {volume} {58}},\
  \bibinfo {pages} {7878} (\bibinfo {year} {1998})}\BibitemShut {NoStop}%
\bibitem [{\citenamefont {Yang}\ and\ \citenamefont
  {Ravi-Chandar}(2001)}]{Yang2001}%
  \BibitemOpen
  \bibfield  {author} {\bibinfo {author} {\bibfnamefont {B.}~\bibnamefont
  {Yang}}\ and\ \bibinfo {author} {\bibfnamefont {K.}~\bibnamefont
  {Ravi-Chandar}},\ }\href {\doibase 10.1016/s0022-5096(00)00022-3} {\bibfield
  {journal} {\bibinfo  {journal} {J. Mech. Phys. Solids.}\ }\textbf {\bibinfo
  {volume} {49}},\ \bibinfo {pages} {91} (\bibinfo {year} {2001})}\BibitemShut
  {NoStop}%
\bibitem [{\citenamefont {Xu}\ \emph {et~al.}(2018)\citenamefont {Xu},
  \citenamefont {Zhang}, \citenamefont {Chen},\ and\ \citenamefont
  {Bobaru}}]{Xu2018}%
  \BibitemOpen
  \bibfield  {author} {\bibinfo {author} {\bibfnamefont {Z.}~\bibnamefont
  {Xu}}, \bibinfo {author} {\bibfnamefont {G.}~\bibnamefont {Zhang}}, \bibinfo
  {author} {\bibfnamefont {Z.}~\bibnamefont {Chen}}, \ and\ \bibinfo {author}
  {\bibfnamefont {F.}~\bibnamefont {Bobaru}},\ }\href {\doibase
  10.1007/s10704-017-0256-5} {\bibfield  {journal} {\bibinfo  {journal} {Int.
  J. Fracture.}\ }\textbf {\bibinfo {volume} {209}},\ \bibinfo {pages} {203}
  (\bibinfo {year} {2018})}\BibitemShut {NoStop}%
\bibitem [{\citenamefont {Sundaram}\ and\ \citenamefont
  {Tippur}(2018)}]{Sundaram2018}%
  \BibitemOpen
  \bibfield  {author} {\bibinfo {author} {\bibfnamefont {B.~M.}\ \bibnamefont
  {Sundaram}}\ and\ \bibinfo {author} {\bibfnamefont {H.~V.}\ \bibnamefont
  {Tippur}},\ }\href {\doibase 10.1016/j.jmps.2018.04.010} {\bibfield
  {journal} {\bibinfo  {journal} {J. Mech. Phys. Solids.}\ }\textbf {\bibinfo
  {volume} {120}},\ \bibinfo {pages} {132} (\bibinfo {year}
  {2018})}\BibitemShut {NoStop}%
\bibitem [{\citenamefont {Cox}(1952)}]{Cox1952}%
  \BibitemOpen
  \bibfield  {author} {\bibinfo {author} {\bibfnamefont {H.~L.}\ \bibnamefont
  {Cox}},\ }\href {\doibase 10.1088/0508-3443/3/3/302} {\bibfield  {journal}
  {\bibinfo  {journal} {Br. J. Appl. Phys.}\ }\textbf {\bibinfo {volume} {3}},\
  \bibinfo {pages} {72} (\bibinfo {year} {1952})}\BibitemShut {NoStop}%
\bibitem [{\citenamefont {McGuigan}\ \emph {et~al.}(2003)\citenamefont
  {McGuigan}, \citenamefont {Briggs}, \citenamefont {Burlakov}, \citenamefont
  {Yanaka},\ and\ \citenamefont {Tsukahara}}]{Mcguigan2003}%
  \BibitemOpen
  \bibfield  {author} {\bibinfo {author} {\bibfnamefont {A.~P.}\ \bibnamefont
  {McGuigan}}, \bibinfo {author} {\bibfnamefont {G.~A.~D.}\ \bibnamefont
  {Briggs}}, \bibinfo {author} {\bibfnamefont {V.~M.}\ \bibnamefont
  {Burlakov}}, \bibinfo {author} {\bibfnamefont {M.}~\bibnamefont {Yanaka}}, \
  and\ \bibinfo {author} {\bibfnamefont {Y.}~\bibnamefont {Tsukahara}},\ }\href
  {\doibase 10.1016/s0040-6090(02)01124-0} {\bibfield  {journal} {\bibinfo
  {journal} {Thin. Solid. Films.}\ }\textbf {\bibinfo {volume} {424}},\
  \bibinfo {pages} {219} (\bibinfo {year} {2003})}\BibitemShut {NoStop}%
\bibitem [{\citenamefont {Cho}\ and\ \citenamefont {Datta}(2019)}]{Jeremy2019}%
  \BibitemOpen
  \bibfield  {author} {\bibinfo {author} {\bibfnamefont {H.~J.}\ \bibnamefont
  {Cho}}\ and\ \bibinfo {author} {\bibfnamefont {S.~S.}\ \bibnamefont
  {Datta}},\ }\href {\doibase 10.1103/PhysRevLett.123.158004} {\bibfield
  {journal} {\bibinfo  {journal} {Phys. Rev. Lett.}\ }\textbf {\bibinfo
  {volume} {123}},\ \bibinfo {pages} {158004} (\bibinfo {year}
  {2019})}\BibitemShut {NoStop}%
\bibitem [{\citenamefont {Bohn}\ \emph {et~al.}(2005)\citenamefont {Bohn},
  \citenamefont {Douady},\ and\ \citenamefont {Couder}}]{Bohn2005}%
  \BibitemOpen
  \bibfield  {author} {\bibinfo {author} {\bibfnamefont {S.}~\bibnamefont
  {Bohn}}, \bibinfo {author} {\bibfnamefont {S.}~\bibnamefont {Douady}}, \ and\
  \bibinfo {author} {\bibfnamefont {Y.}~\bibnamefont {Couder}},\ }\href
  {\doibase 10.1103/PhysRevLett.94.054503} {\bibfield  {journal} {\bibinfo
  {journal} {Phys. Rev. Lett.}\ }\textbf {\bibinfo {volume} {94}},\ \bibinfo
  {pages} {054503} (\bibinfo {year} {2005})}\BibitemShut {NoStop}%
\bibitem [{\citenamefont {Song}\ \emph {et~al.}(2010)\citenamefont {Song},
  \citenamefont {Meng}, \citenamefont {Xu},\ and\ \citenamefont
  {Shao}}]{Song2010}%
  \BibitemOpen
  \bibfield  {author} {\bibinfo {author} {\bibfnamefont {F.}~\bibnamefont
  {Song}}, \bibinfo {author} {\bibfnamefont {S.}~\bibnamefont {Meng}}, \bibinfo
  {author} {\bibfnamefont {X.}~\bibnamefont {Xu}}, \ and\ \bibinfo {author}
  {\bibfnamefont {Y.}~\bibnamefont {Shao}},\ }\href {\doibase
  10.1103/PhysRevLett.104.125502} {\bibfield  {journal} {\bibinfo  {journal}
  {Phys. Rev. Lett.}\ }\textbf {\bibinfo {volume} {104}},\ \bibinfo {pages}
  {125502} (\bibinfo {year} {2010})}\BibitemShut {NoStop}%
\bibitem [{\citenamefont {Aydin}\ and\ \citenamefont
  {Degraff}(1988)}]{Aydin1988}%
  \BibitemOpen
  \bibfield  {author} {\bibinfo {author} {\bibfnamefont {A.}~\bibnamefont
  {Aydin}}\ and\ \bibinfo {author} {\bibfnamefont {J.~M.}\ \bibnamefont
  {Degraff}},\ }\href {\doibase 10.1126/science.239.4839.471} {\bibfield
  {journal} {\bibinfo  {journal} {Science.}\ }\textbf {\bibinfo {volume}
  {239}},\ \bibinfo {pages} {471} (\bibinfo {year} {1988})}\BibitemShut
  {NoStop}%
\bibitem [{\citenamefont {Ghabache}\ \emph {et~al.}(2016)\citenamefont
  {Ghabache}, \citenamefont {Josserand},\ and\ \citenamefont
  {S\'eon}}]{Ghabache2016}%
  \BibitemOpen
  \bibfield  {author} {\bibinfo {author} {\bibfnamefont {E.}~\bibnamefont
  {Ghabache}}, \bibinfo {author} {\bibfnamefont {C.}~\bibnamefont {Josserand}},
  \ and\ \bibinfo {author} {\bibfnamefont {T.}~\bibnamefont {S\'eon}},\ }\href
  {\doibase 10.1103/PhysRevLett.117.074501} {\bibfield  {journal} {\bibinfo
  {journal} {Phys. Rev. Lett.}\ }\textbf {\bibinfo {volume} {117}},\ \bibinfo
  {pages} {074501} (\bibinfo {year} {2016})}\BibitemShut {NoStop}%
\bibitem [{\citenamefont {Vandenberghe}\ \emph {et~al.}(2013)\citenamefont
  {Vandenberghe}, \citenamefont {Vermorel},\ and\ \citenamefont
  {Villermaux}}]{Vandenberghe2013}%
  \BibitemOpen
  \bibfield  {author} {\bibinfo {author} {\bibfnamefont {N.}~\bibnamefont
  {Vandenberghe}}, \bibinfo {author} {\bibfnamefont {R.}~\bibnamefont
  {Vermorel}}, \ and\ \bibinfo {author} {\bibfnamefont {E.}~\bibnamefont
  {Villermaux}},\ }\href {\doibase 10.1103/PhysRevLett.110.174302} {\bibfield
  {journal} {\bibinfo  {journal} {Phys. Rev. Lett.}\ }\textbf {\bibinfo
  {volume} {110}},\ \bibinfo {pages} {174302} (\bibinfo {year}
  {2013})}\BibitemShut {NoStop}%
\bibitem [{\citenamefont {Baumberger}\ \emph {et~al.}(2008)\citenamefont
  {Baumberger}, \citenamefont {Caroli}, \citenamefont {Martina},\ and\
  \citenamefont {Ronsin}}]{Baumberger2008}%
  \BibitemOpen
  \bibfield  {author} {\bibinfo {author} {\bibfnamefont {T.}~\bibnamefont
  {Baumberger}}, \bibinfo {author} {\bibfnamefont {C.}~\bibnamefont {Caroli}},
  \bibinfo {author} {\bibfnamefont {D.}~\bibnamefont {Martina}}, \ and\
  \bibinfo {author} {\bibfnamefont {O.}~\bibnamefont {Ronsin}},\ }\href
  {\doibase 10.1103/PhysRevLett.100.178303} {\bibfield  {journal} {\bibinfo
  {journal} {Phys. Rev. Lett.}\ }\textbf {\bibinfo {volume} {100}},\ \bibinfo
  {pages} {178303} (\bibinfo {year} {2008})}\BibitemShut {NoStop}%
\bibitem [{\citenamefont {Kolvin}\ \emph {et~al.}(2015)\citenamefont {Kolvin},
  \citenamefont {Cohen},\ and\ \citenamefont {Fineberg}}]{Kolvin2015}%
  \BibitemOpen
  \bibfield  {author} {\bibinfo {author} {\bibfnamefont {I.}~\bibnamefont
  {Kolvin}}, \bibinfo {author} {\bibfnamefont {G.}~\bibnamefont {Cohen}}, \
  and\ \bibinfo {author} {\bibfnamefont {J.}~\bibnamefont {Fineberg}},\ }\href
  {\doibase 10.1103/PhysRevLett.114.175501} {\bibfield  {journal} {\bibinfo
  {journal} {Phys. Rev. Lett.}\ }\textbf {\bibinfo {volume} {114}},\ \bibinfo
  {pages} {175501} (\bibinfo {year} {2015})}\BibitemShut {NoStop}%
\bibitem [{\citenamefont {Kolvin}\ \emph {et~al.}(2017)\citenamefont {Kolvin},
  \citenamefont {Fineberg},\ and\ \citenamefont {Adda-Bedia}}]{Kolvin2017a}%
  \BibitemOpen
  \bibfield  {author} {\bibinfo {author} {\bibfnamefont {I.}~\bibnamefont
  {Kolvin}}, \bibinfo {author} {\bibfnamefont {J.}~\bibnamefont {Fineberg}}, \
  and\ \bibinfo {author} {\bibfnamefont {M.}~\bibnamefont {Adda-Bedia}},\
  }\href {\doibase 10.1103/PhysRevLett.119.215505} {\bibfield  {journal}
  {\bibinfo  {journal} {Phys. Rev. Lett.}\ }\textbf {\bibinfo {volume} {119}},\
  \bibinfo {pages} {215505} (\bibinfo {year} {2017})}\BibitemShut {NoStop}%
\bibitem [{\citenamefont {Kolvin}\ \emph {et~al.}(2018)\citenamefont {Kolvin},
  \citenamefont {Cohen},\ and\ \citenamefont {Fineberg}}]{Kolvin2017b}%
  \BibitemOpen
  \bibfield  {author} {\bibinfo {author} {\bibfnamefont {I.}~\bibnamefont
  {Kolvin}}, \bibinfo {author} {\bibfnamefont {G.}~\bibnamefont {Cohen}}, \
  and\ \bibinfo {author} {\bibfnamefont {J.}~\bibnamefont {Fineberg}},\ }\href
  {\doibase 10.1038/NMAT5008} {\bibfield  {journal} {\bibinfo  {journal} {Nat.
  Mater.}\ }\textbf {\bibinfo {volume} {17}},\ \bibinfo {pages} {140} (\bibinfo
  {year} {2018})}\BibitemShut {NoStop}%
\end{thebibliography}
%

\end{document}